\documentclass[11pt]{article}
\usepackage[utf8]{inputenc}
\usepackage{authblk}
\usepackage{natbib}
\usepackage{babel}
\usepackage{graphicx}
\usepackage{adjustbox}
\usepackage{tabularx}
\usepackage{geometry}
\usepackage{setspace} 
\usepackage{caption}
\usepackage{subcaption}
\usepackage{hyperref}
\hypersetup{
	colorlinks,
	linkcolor={red!50!black},
	citecolor={blue!50!black},
	urlcolor={blue!80!black}
}
\usepackage{amsmath,amssymb}
\usepackage{bm}
\usepackage{longtable}
\usepackage{multicol}
\usepackage{multirow}
\usepackage{booktabs}
\usepackage{makecell}
\usepackage{comment}
\usepackage{xcolor}
\usepackage{soul}
\usepackage{tikz}
\usepackage{pgfplots}
\pgfplotsset{compat=1.17} 
\usepackage{array}
\usepackage{float}

\title{Environmental migration? \\ \large An overview of the literature.}
\author[1]{Maria Cipollina}
\author[2,3]{Luca De Benedictis}
\author[2]{Elisa Scibè\thanks{\scriptsize Corresponding author. The authors would like to acknowledge the contribution of MUR, which partially funded the visit of Elisa Scibè to the Universitè Libre de Bruxelles, Belgium. Many thanks to Francesca Pallotti, Tom Stanley, the Editor Galina Hale and the anonymous referee of the Review and the participants to various seminars and conferences (EUSN2021; MAER-Net Colloquim 2021) for the insightful comments.  No conflict of interest applies.}}

\affil[1]{\small University of Molise - cipollina@unimol.it}
\affil[2]{\small University of Macerata - luca.debenedictis@unimc.it; e.scibe1@unimc.it}
\affil[3]{\small Luiss University}

\date{March 30, 2024}

\begin{document}

\begin{singlespacing}
\maketitle
\vspace{-10mm}

\begin{abstract}

\noindent 
This article provides a comprehensive quantitative overview of the literature on the relation-ship between environmental changes and human migration. It begins with a systematic approach to bibliographic research and offers a bibliometric analysis of the empirical contribu-tions. Specifically, we map the literature and conduct systematic research using main bibliographic databases, reviews, and bibliometric analysis of all resulting papers. By constructing a citation-based network, we identify four separate clusters of papers grouped according to certain characteristics of the analysis and resulting outcomes. Finally, we apply a meta-analysis to a sample of 96 published and unpublished studies between 2003 and 2020, providing 3,904 point estimates of the effect of slow-onset events and 2,065 point estimates of the effect of fast-onset events. 

Overall, the meta-analytic average effect on migration is small for both slow- and rapid-onset events; however, it is positive and significant. Accounting for the clustering of the literature, which highlights how specific common features of the collected studies influence the magnitude of the estimated effect, reveals a significant heterogeneity among the four clusters of papers. This heterogeneity gives rise to new evidence on the formation of club-like convergence of literature outcomes.
\end{abstract}


\end{singlespacing}

\section{Introduction}
In a world of changing climate and increasing occurrence of natural hazards, the role of environ- mental factors in shaping migration patterns has become a most debated topic within institutions and academia. As opposed to a simplistic vision of a general direct role of environmental factors in determining migration flows from environmentally stressed areas and regions hit by calamities, more complex scenarios have emerged, with analyses reporting different and sometimes opposite outcomes. This may not only be due to the intrinsic complexity of their extent and scale, but also to differences in specific characteristics of scientific contributions \citep{IOM2021}. 

The literature on the relationship between environmental factors and human mobility is characterized by heterogeneous findings: some contributions highlight the role of climate changes as a driver of migratory flows, while others underline how this impact is mediated by geographical, economic, and the features of the environmental shock.  
This paper aims to map the economic literature on these topics moving away from a classical literature review and offering a methodology that integrates three approaches in a sequence, in this way we believe that our contribution improves the existing literature on several dimensions. First, the analysis starts with systematic research of the literature through main bibliographic databases and collecting previous reviews and meta-analyses, followed by a review and bibliometric analysis of all resulting papers. This step produces a sample of 151 papers empirical and non-empirical contributions, spanning the last 20 years and focusing on different geographical areas, taking into account different socio-economic factors, applying different methodologies and empirical approaches to the analysis of slow-onset climatic events and/or fast-onset natural catastrophic events. Most importantly, the sample provides a variety of different outcomes on the impact of climatic changes and hazards on migration, revealing three main possible scenarios: (1) active role of environmental factors as a driver of migration; (2) environmental factors as a constraint to mobility; (3) non-significant role of environmental factors among other drivers of migration. 

Second, to investigate the determinants of this extreme heterogeneity of outcomes, we postulate the assumption that the inter-connectivity of papers may play a role in shaping such different conclusions. Considering the ensemble of papers referenced by each contribution included in the sample, as a second step, we build a bibliographic coupling network, where papers are linked to each other according to the number of shared references. This citation-based method allows for the formation of a network of contributions in the literature space and highlights some potential common grounds among papers. We then run a community detection of the resulting network that produces four main clusters that gather papers together according to not only certain characteristics of the analysis but also resulting outcomes.

Finally, we use the clustered structure in the last step of the analysis: a Meta-Analysis (MA) to summarize and analyze all estimated effects of environmental variables on human mobility. The MA is a ``quantitative survey" of  empirical economic evidence on a given hypothesis, phenomenon, or effect, and provides a statistical synthesis of results from a series of studies \citep{Stanley2001}. The MA can be applied to any set of data and the synthesis will be meaningful only if the studies have been collected systematically \citep{Borenstein2009}. A highly significant result can be potentially considered as a consensual indication of the external validity of the correlation of the phenomena under scrutiny.

Therefore, from the original 151 paper we build - through a replicable process of
screening, eligibility, and inclusion of contribution based on PRISMA guideline (see Figure 1 in the next section) - a unique dataset that synthesises the estimated coefficients of 96 empirical papers released between 2003 and 2020, published in academic journals, working papers series, or unpublished studies, providing 3,904 point estimates of the effect of slow-onset events (e.g. climate change) and 2,065 point estimates of the effect of fast-onset natural events(e.g. catastro-phes) on different kinds of human mobility (international, domestic, and with a clear pro-urban directionality). Overall, the meta-analytic average effect estimates a small impact of slow- and rapid-onset variables on migration, however positive and significant. When the communities of papers are accounted for, however, a significant heterogeneity emerges among the four clusters of papers, giving rise to new evidence on the limits of a consensual effect of climatic shocks on permanent human displacement and the formation of club-like convergence of literature outcomes.

This is not the first MA on environmental migration. Concerning previous published reviews \citep{hoffmann2020meta, Sedova_2020, beine_jeusette_2021, Hoffmann_2021} our article contributes and adds to the existing literature: (a) providing systematic research of the literature through main bibliographic databases, followed by a review and bibliometric analysis of all resulting papers; (b) building a citation-based network of contribu-tions, that allows identifying four separate clusters of papers; (c) applying MA methods on a much larger sample of both micro- and macro-level estimates of environmental factors (slow- and fast-onset events) as a driver of migration (international and internal, including urbanization).
Moreover, our overview highlights the role of the interconnectivity of studies in driving some main findings of the environmental migration literature.

Section \ref{sec: review} offers a systematic review of the literature and gives a detailed description of the data collection process; Section \ref{sec: network} analyses the structural characteristic of the network of the bibliographically coupled papers; Section \ref{sec: meta} summarizes and discusses the results of the MA, finally, Section \ref{sec: conclusions} concludes and offers some possible future extensions of the analysis.

\section{Systematic review}
\label{sec: review}

This section reports the different phases of the systematic review. We do it schematically to facilitate the understanding of the proposed procedure.

- \textit{Setting the boundaries of the literature.} 
This first step provides the most comprehensive sample of economic contributions on the relationship between climatic variations (and natural hazards) and human mobility, in all its different forms. We implement a systematic review aimed at mapping the body of literature and defining the boundaries of our focus. Systematic reviews have become highly recommended to conduct bibliographic overviews of specific literature because they provide a tool to report a synthesis of the state of the art of a field through a structured and transparent methodology \citep{page2021prisma}. To allow for comparability with previous MA and reviews, we also add to our sample all articles included in two recently published MA, \cite{hoffmann2020meta} and \cite{beine_jeusette_2021}\footnote{\ A detailed table highlighting specific studies featured in other meta-analyses, along with their citations, that have been reviewed in our study is provided in the Supplementary material, Section A.}. We begin with the definition of the research question and the main keywords, to gather and collect data in a sample of contributions. After the definition of inclusion and exclusion conditions, we proceed with a screening by title to exclude off-topic contributions and then to a screening of the text to assure the uniformity of contributions. The resulting sample is then the object of a preliminary bibliometric analysis.

- \textit{Defining the research question and keywords.} The purpose of our systematic search is to collect all possible economic contributions to the impact of environmental factors on migration determinants. We define three keywords of the three phenomena under analysis: 
- climate change, as the most investigated environmental factor in the literature. The events connected to climate change are hereby intended as slow-onset events that gradually modify climatic conditions in the long run. We specifically focus on variations of temperature, precipitation, and soil quality (such as desertification, salinity, or erosion), factors that are not expected to cause an immediate and sudden expected impact, but slowly modify environmental conditions;
- natural disasters, defined as fast-onset events that introduce a sudden shock (see Appendix Table 1);
- migration, which captures all possible patterns of human mobility, including within the borders of a country, which might be a potential response to environmental change. Most importantly, internal mobility includes also the process of urbanization of people moving out of rural areas to settle in cities.

- \textit{Collecting data and initial search results.} To collect data we use two main literature databases, namely Scopus and Web of Science.\footnote{\ The extraction is made through \textsf{bibliometrix}, an R tool for science mapping analysis that reads and elaborates the information exported by Scopus and Web of Science \citep{Aria2017959}.} Exploiting the specific indexing and keyword definition of both sources, the search is run allowing for any kind of document type (articles in journals, book chapters, etc.) but limiting the area to economic literature in English.\footnote{\ Scopus: \texttt{key}(``migration" \texttt{and} (``natural disasters" \texttt{or} ``climate change")) \texttt{and} (\texttt{limit-to} (\texttt{subjarea},``econ")) \texttt{and} (\texttt{limit-to}(\texttt{language},``English")), Date: 24/11/2020. Web of Science: ((\texttt{AK}=(migration  \texttt{and} (``natural disasters"  \texttt{or} ``climate change")))  \texttt{or} (\texttt{KP} = (migration  \texttt{and} (``natural disasters"  \texttt{or} ``climate change"))))  \texttt{and language}: (English) Refined by: \texttt{web of science categories}: (``economics"), Date: 24/11/2020.} 
The obtained sample only includes published documents, however since we perform a MA, it is important to take into account also non-published documents, as a way to control for a well-known publication bias in meta-analytic methodology (see Section \ref{sec: meta}). Therefore, we use the bibliographic database IDEAS, based on RePEc and dedicated to Economics, to include unpublished and working papers.\footnote{\ We use the Advanced Search tool, searching by Keywords and Title: migration \texttt{and} (``natural disasters" \texttt{or} ``climate change").} A selection of the contributions is made manually. Finally, to meet the purpose of comparability with other recent meta-analyses on the impact of environmen-tal factors on migration, we also include all the contributions that have been reviewed in two main articles: \cite{hoffmann2020meta} that provide a MA on 30 empirical papers focusing on country-level studies and \cite{beine_jeusette_2021} that review 51 papers and offer an investigation of the role of methodological choices of empirical studies (at any level) on the sign and magnitude of estimated results. Merging the results gives a sample of 203 records.

- \textit{Screening of the results.} We manually and meticulously screen the collected items through Scopus and Web of Science by title and we exclude papers on the migration of animals, plants, or other species, or focusing on topics different from human mobility (i.e. discrimination, crime, wars) or on the impact of environmental variables not corresponding to our definition of environmental factors (air pollution, mineral resources). All the papers in \cite{beine_jeusette_2021}, \cite{hoffmann2020meta} and those manually selected from IDEAS RePEc are automatically included in the sample with no concern of incoherence.
The screening by title leads to the exclusion of 20 papers. The remaining 183 documents underwent a text screening process, which involved a careful and thorough reading of each paper to isolate eligible content. This stage leads to the removal of additional 32 documents covering on the one hand the analysis of the impact of environmental variables at destination countries (thus not focusing on their role on migration determinants at origin). We also exclude all the papers in which the dependent variable of the empirical exercise is not a measure of human mobility (i.e. remittances, poverty, wealth, employment, etc.). After duplicates removal, the sample results in 151 documents of different kinds: 35 records are non-empirical and contain an ensemble of literature reviews, qualitative analysis, theoretical modeling, and policy papers; 116 records are categorized as empirical, in which the dependent variable is a measure of human mobility and at least one environmental variable is an independent variable. 

\begin{figure}[ht]
\caption{PRISMA Diagram}
\begin{center}
\includegraphics[scale=0.75]{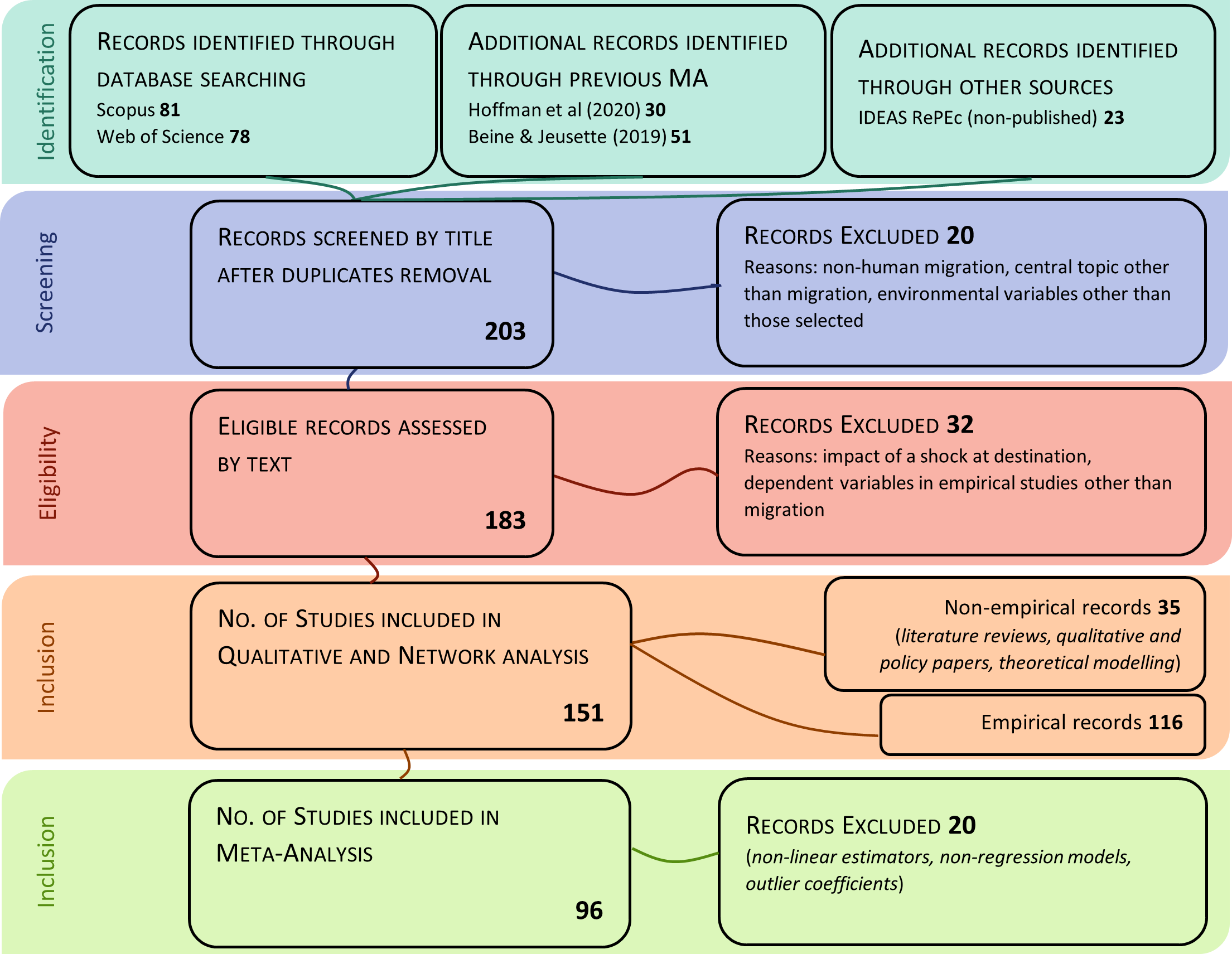}
\label{fig:prisma}
\end{center}
{\scriptsize {\bf Note}: {\linespread{0.5}\selectfont PRISMA Diagram \citep{prisma2020} of identification, screening, eligibility and inclusion stages of academic contributions. The resulting sample is obtained through a search on Scopus, Web of Science, Google Scholar, IDEAS RePEc, and previous meta-analyses \citep{hoffmann2020meta, beine_jeusette_2021}. \par}}
\end{figure}

The PRISMA flow diagram \citep{Moherb} in  Figure \ref{fig:prisma} shows the process of identifica-tion, screening, eligibility, and inclusion of contributions in the final sample. It is important to note that there are two levels of inclusion: the first level identifies the sample of contributions included in our network analysis, while the second level is restricted to quantitative analyses suitable for the MA. To conduct a MA it is crucial to select only comparable papers that provide complete information (mainly on estimated coefficients and standard errors) that can then be used to recover the average effect size \footnote{\ Our inclusion criteria prioritize studies reporting outcomes in an appropriate and consistent manner. In particular, we have excluded studies that do not rely on a complete set of objective measures. For instance, studies that only present estimated coefficients, solely indicating the significant level, without reporting standard errors or $t$-ratios have been excluded because they do not allow for the calculation of a meta-synthesis.}. This implies the exclusion of papers that do not comply with the requirements of a MA. However, those excluded papers can be of interest in building the taxonomy of the whole concerned literature, as they may play a role in building links between different contributions (see Section \ref{sec: network}). Similarly, non-quantitative (policy, qualitative or theoretical) papers may participate as well in the development of research fronts or give a direction to a certain thread of contributions and incidentally affect the detection of clusters. These reasons led us to build our citation-based network and perform the network analysis and the community detection on the whole sample, while only the sample for the MA is restricted only to quantitative contributions that meet the coding requirements. Our final database of point estimates for the MA includes 96 papers released between 2003 and 2020, published in an academic journal, working papers series, or unpublished studies, providing 3,904 point estimates of the effect of slow-onset events (provided by 66 studies) and 2,065 point estimates of the effect of fast-onset events (provided by 60 studies). The list of articles is in the Appendix Table 2.

\subsection{Bibliometric Analysis}
This section summarizes the most relevant features of the ensemble of economic literature collected in our sample.\footnote{\ All records have been uploaded and summary statistics produced using the R tool \textsf{bibliometrix} \citep{Aria2017959}. Scopus and Web of Science allow for the download in the bulk of records in \textsf{.bibtex} format, ready to be converted in R objects. Other records are manually entered, depending on the publication status of the single record: for published documents additional research of the specific document is made on Scopus and the relative \textsf{.bibtex} file is downloaded and added to the other results; for unpublished papers, which cannot be found in the two sources, a \textsf{.bibtex} is manually created following the structure of fields and information in the downloaded ready-to-use files. After merging each file and removing duplicates we obtain the data source that contains the bibliographic information of all articles, including publication year/latest draft, author(s), title, journal, keywords, affiliations, and references.}

\begin{figure}[t]
\caption{Number of documents per year}
   \begin{center}
  \includegraphics[scale=0.5]{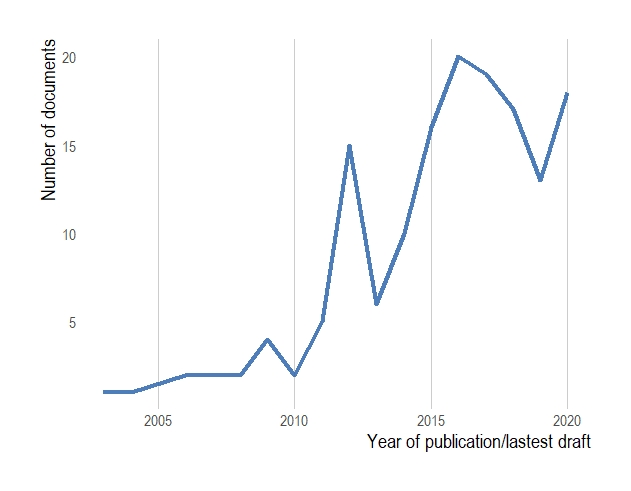}
    \end{center}
    \label{fig:yearpublication}
    {\scriptsize {\bf Note}: {\linespread{0.5}\selectfont Sample of academic contributions about migration and environmental factors from Scopus, Web of Science, Google Scholar, IDEAS RePEc, and previous meta-analyses \citep{hoffmann2020meta, beine_jeusette_2021} collected, merged, screened and included by the authors. \par}}
\end{figure}

The economic literature started to pay attention to the potential relevance of environmental events on migration in the early 2000s, although the topic had already gained some relevance in global debate decades before, and scientific production increased sharply in the last 17 years. 
Figure \ref{fig:yearpublication} shows that the scientific production in the specific field is quite recent, spanning from 2003 to 2020, with a peak of 20 contributions in 2016 and an annual growth rate for the overall period at 18.5 percent. Taking a closer look at the cited references, it is possible to trace back an article published before 2003 \citep{Findley1994539}, that provides a qualitative analysis of drought-induced mobility in Mali (finding no evidence of any role of 1983-85 droughts on migration).
As our research of documents is based on keywords, naturally the three most repeated are those put in the search key (``migration", ``climate change" and ``natural disasters").\footnote{\ Variants of words or concepts have been aggregated in a unique item i.e. \textit{climate change} and \textit{climatic change} or \textit{environmental migrants} and \textit{environmental migration}.} Within the topic of migration, there's a greater emphasis on international mobility compared to internal migration.  However, internal migration may include also urbanization or rural-urban migration,  and when combined, they are as common as international migration (counting 21 repetitions per group). Environmental migration is also explored as a form of \textit{forced migration}, originating refugees, or specifically environmental refugees. The keywords related to environmental issues are more focused on slow-onset events like (\textit{rainfall, temperature, global warming} and \textit{climate variability}) rather than rapid-onset events. Although, some of the latter are more recurrent than others, such as \textit{drought, floods} and ultimately \textit{earthquakes}. 

\begin{figure}[t]
    \caption{The 20 most relevant publication sources by field}
    \begin{center}
    \includegraphics[scale = 0.75]{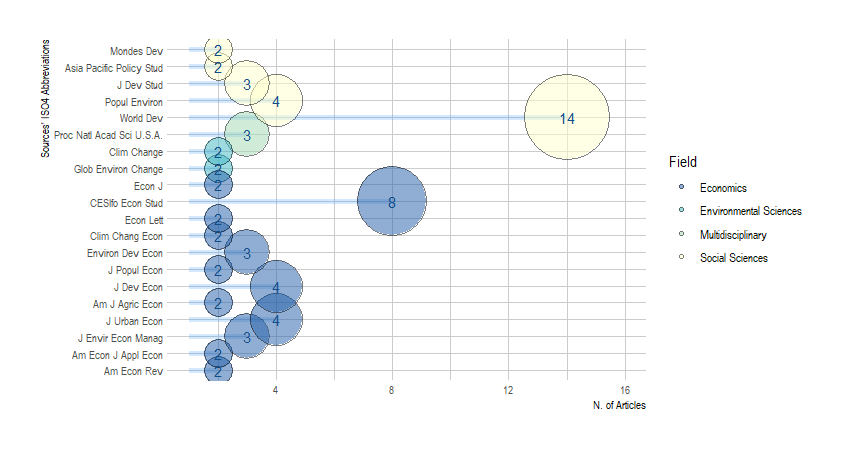}
    \end{center}
    \label{fig:soumostrel}
    {\scriptsize {\bf Note}: {\linespread{0.5}\selectfont Sample of academic contributions about migration and environmental factors from Scopus, Web of Science, Google Scholar, IDEAS RePEc, and previous meta-analyses \citep{hoffmann2020meta, beine_jeusette_2021} collected, merged, screened and included by the authors. \par}}
\end{figure}

Overall 288 authors have contributed to this literature, with 372 appearances, 
34 documents are single-authored, the mean number of authors per document is 1.88; when considering exclusively multi-authored documents, the number of co-authors per document rises to 2.16, with a maximum of co-authors of 9.
Various disciplines have put attention to the topic. Despite journals specializing in economics and econometrics representing the majority of the sources of publication, the literature includes also other disciplines (Figure \ref{fig:soumostrel}).
 
Specifically, economic environmental migration is the object of publication in journals specialized in environmental sciences, geography, and social sciences such as urban studies, agriculture, demography, and political studies. A special mention has to be done for development studies: many reviews and journals specialized in development have issued contributions on the topic, highlighting the trend of observing the topic through development lenses. As an example, 14 documents in our sample are published in \textit{World Development}, a multi-disciplinary journal of development studies.

A picture of the most relevant documents included in the sample is provided by simple measures, such as the number of global citations as reported in Scopus (at the moment of the bulk download of all sources), and the number of local citations, which shows how many times a document has been cited by other papers included in the sample. Measures for the most cited documents (global and local citation scores) in the sample are reported in Appendix Table 3. The difference between global and local citation scores (almost four times higher) reveals that the documents have been cited by papers not included in our sample. It means that environmental migration has attracted the interest of different disciplines or they became part of the two main strands of literature, climate change, and migration, separately. 58 papers have not been cited in any of our samples, while 52 have zero citations globally. A part of it can be explained by the 18 papers that have been published recently in 2020, which could not have been cited yet because of timing (except for some contributions published in early 2020 such as \cite{Mueller2020} and \cite{Rao2020}.\footnote{\ The issue of timing will be addressed in the network analysis, choosing a specific type of citation-based network, the bibliographic coupling network, to minimize the risk of missing connections between papers.} Position and the number of citations confirm the central role of papers published by Gray Clark and Valerie Mueller \citep{Gray2012pnas, Gray2012worlddev, Mueller2014}, receiving high citations both globally and internally. Some papers seem to be more relevant locally than globally: \cite{Marchiori2012} and \cite{Beine2015} had a bigger influence on our sample of economic environmental migration literature rather than globally, scoring the highest number of local citations. Conversely, \cite{Hornbeck2012} seems to be cited more in literature outside the specific literature of environmental migration. 

\subsection{Overview of major results}

The literature on the effects of climate and natural disasters on migration is characterized by a rich variety of studies both in micro- and macro-economic analyses. Country-level analyses tend to find evidence of a direct or indirect impact of environmental factors on migration patterns, either internally or internationally. \cite{Barrios2006} and \cite{Marchiori2012} find evidence of an increase in internal migration, especially towards urban areas in the case of Sub-Saharan Africa, according to many specific historical and developmental factors. Both contributions highlight how worsening climatic conditions correspond to a faster urbanization process. \cite{Marchiori2012} add also that this climate-driven urbanization process results also in higher international migration rates, acting as a channel of transmission of the effect of climate. 

The macro literature, in line with most validated theoretical models of migration, also investigates whether the effect is conditioned to income levels of the country of origin of potential migrants \citep{Marchiori2012, Beine2015, Beine2017}. The role of income in a specific origin country experiencing the effects of environmental events is found to be crucial to determine the sign and the magnitude of the impact. \cite{Cattaneo2016} support from one side the active role of those events in fostering migration, but show how this effect is conditioned to middle-income countries. The effect is the opposite when conditioning the analysis to poor countries, highlighting the existence of certain \textit{constraints} to mobility. Worsened environmental conditions may exacerbate liquidity constraints or lack of access to credit aimed at financing the migratory project, which lead to what has been called \textit{poverty trap}. Furthermore, these conditioned results seem to be robust even when another important channel is controlled, agricultural productivity. Climatic conditions and disruptive hazards may constitute major drawbacks for agricultural productivity, leading the agriculture-dependent part of the population to move out from rural areas: \cite{Cai2016} and \cite{Coniglio2015} provide evidence of an indirect link between worsened temperature and precipitation conditions and migration, mediated by the level of agricultural dependency of the country of origin. Sudden and fast-onset hazards, on the other side, are not found to contribute significantly to human mobility, except in the case of a higher-educated population, more mobile than other groups after the disruption of a natural disaster \citep{Drabo2015}.

\begin{figure}[t]
    \caption{Number of case studies covered by the micro-level sub-sample per country}
    \label{fig:casestudies}
    \begin{center}
    \includegraphics[scale=0.75]{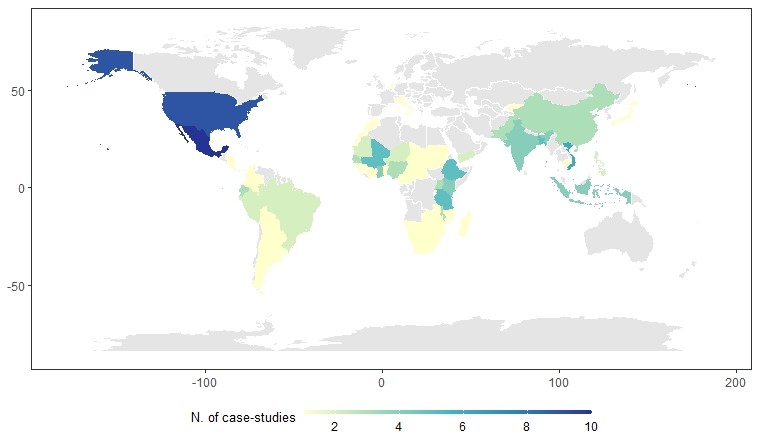}
    \end{center}
    {\scriptsize {\bf Note}: {\linespread{0.75}\selectfont Sub-sample of micro-level studies about migration and environmental factors from Scopus, Web of Science, Google Scholar, IDEAS RePEc, and previous meta-analyses \citep{hoffmann2020meta, beine_jeusette_2021} collected, merged, screened and included by the authors. \par}}
\end{figure}
 
Micro-level literature provides a vast variety of case studies on different potential impacts of environmental factors on mobility. In our sample, they almost double macro-level contributions (86 contributions against 47) and provide different scenarios. Firstly, while macro-level studies mostly provide analyses at the global level or for some groups of countries or macro-regions, micro-level analyses tend to observe a specific phenomenon hitting a specific area or to study differences in the impact of a common phenomenon in different areas. The most covered region as a whole is Sub-Saharan Africa, with 65 case studies included in the contributions (Figure \ref{fig:casestudies}).\footnote{\ Some contributions are not single-case studies.} When the level of analysis is less aggregated than the national or sub-national level, and individual or household behavior is observed through the use of surveys, the picture gains complexity and less generalized conclusions. This seems clear in \cite{Gray2016} who analyze a series of comparable surveys across five Sub-Saharan countries, which have consistent differences. The heterogeneity of responses to climatic variations across those countries is strictly linked to the characteristics of the area and of the specific households. Poorer countries (such as Burkina Faso) mainly experience internal and temporary migration, often on a rural-rural channel as a way to diversify risk \citep{Henry2003, Henry2004}. Long-distance migration seems to be constrained by liquidity and access to credit to finance those expensive journeys. Migratory trends of Nigerian households are pushed in times of favorable climatic conditions, while the effect of adverse conditions interacts with a negative effect on income and traps populations at origin \citep{Cattaneo2019Clim}. Overall, micro-level studies focused on the African continent highlight the importance of considering the interplay of a variety of factors when it comes to the analysis of the role of environmental factors, defining the new path toward hybrid literature.

The single countries that receive singularly the most attention are Mexico, with 10 case studies, and the U.S., with 9 case studies. This should not be a surprise because of two reasons: firstly, the stock of Mexican emigrates has been constantly the highest in the world (in absolute terms) as well as the migratory flow between Mexico and the U.S. But there might also be a publication-related reason based on the fact that the vast majority of journals in our sample are U.S. based. Major findings support the relevance of environmental drivers (mainly precipitation shortage) on push factors from Mexico (\cite{Feng2010} estimates that a 10\% reduction of agricultural productivity driven by scarce rainfall corresponds to the rise of 2\% of emigrants).

Southern and Eastern Asia, representing by far the most disaster-prone area in the world.\footnote{\ Asia suffered the highest number of disaster events. In total, between 2000 and 2019,
there were 3,068 disaster events in Asia, followed by the 1,756 events in the Americas and 1,192 events in Africa \citep{UNDRR}.} also provide a variety of heterogeneous scenarios. The case of Vietnam \citep{Koubi2016worlddev, Berlemann2020} shows how the Vietnamese population chooses different coping strategies in response to different kinds of environmental stressors. While gradual climatic variations lead to mechanisms of adaptation \textit{in loco} to new climatic conditions, sudden shocks drive the decision to migrate elsewhere. However, mobility responses to different types of hazards might be different according to their specific consequences and duration \citep{Berlemann2020}. On the contrary, the case of Bangladesh supports the hypothesis that the existence of previous barriers to access to migration is worsened by the occurrence of disasters, specifically in the face of recurrent and intense flooding \citep{Gray2012pnas}.

The specific case of earthquakes across the world (El Salvador in \cite{Halliday2006}, Japan in \cite{Kawawaki2018} and Indonesia in \cite{Gignoux2016} for instance) shows a common trend of outcomes: highly disruptive disasters such as earthquakes tend to decrease mobility from the hit area. An interesting mechanism to explain this common trend found in three very different contexts is given by, not only the already mentioned financial constraints but also the possibility of higher local employment opportunities due to post-disaster reconstruction \citep{Gignoux2016, Halliday2006}. Moreover, households are found to respond to hazard by using the labor force as a buffer to the damages and redistributing labor within the household, with female mobility drastically dropping more than males and being substituted with increased hours of domestic labor \citep{Halliday2012}.

Analyses on South American countries also contribute to giving a hint of the complexity of the phenomenon. \cite{Thiede2016} show how internal migration is indeed impacted by rising temperature when considering the general effect; however, it hides an extreme heterogeneity of outcomes when specific characteristics of the areas and individuals are taken into account, resulting in a non-uniform effect.

An evident gap in the literature emerges in Figure \ref{fig:casestudies}: European countries have rarely been the object of study of the impact of environmental factors on mobility. This might be motivated by the fact that the European continent is mostly seen as a destination for migrants than an origin. It should not surprise that the two articles covering European countries, namely Italy \citep{Spitzer2020} and the Netherlands \citep{Jennings2015} analyze historical data of mobility at the beginning of the XX century (respectively earthquake in Sicily and Calabria and climate variability associated with riverine flooding in the Netherlands). Nevertheless, figures show that Europe is not unrelated to the occurrence and frequency of hazards as well as to sizable internal mobility that should receive some attention.

\section{The inter-connectivity of papers}
\label{sec: network}

Our quantitative approach aims at analyzing the connectivity that exists among papers according to a citation-based approach and detecting the existence of communities or clusters. Since our target literature is characterized by a high heterogeneity of results, both in the direction and magnitude of the impact, we try to investigate the existence of potential specific patterns that lead to a certain type of analysis, methodology, or result under network-analysis lenses. We then use all information from this section to implement the meta-analysis. 

\subsection{Bibliographic coupling and citation-based approaches}

The citation-based approach we choose is called bibliographic coupling.\footnote{\ Bibliographic coupling, first introduced by \citep{Kessler196349, kessler1963bibliographic}, belongs to the broader class of citation-based approaches to science mapping.
Co-citation is based on the relationship established by citing authors of a paper: two papers are linked whenever they jointly appear in the cited references of at least a third paper. Direct citation is the most intuitive approach, linking two papers if one has cited a precedent one. As co-citation, direct citation performs better for long time windows to visualize historical connections \citep{Klavans2017984}. In terms of accuracy, it has been established that direct citation provides a more accurate representation of the taxonomy of scientific production \citep{Klavans2017984}, but for the specific requirements the methodology imposes, it has not gained much success \citep{Boyack20102389}.} Two scientific papers ``bear a meaningful relation to each other when they have one or more references in common". Thus, the fundamentals of the link between two papers are depicted by the number of shared papers they both include in their references, which constitute the strength of the connectivity they have. In other words, a reference that is cited by two papers constitutes a ``unit of coupling between them" \citep{kessler1963bibliographic}. Two articles are then said bibliographically coupled if at least one cited source appears in both articles \citep{Aria2017959}. Bibliographic coupling is increasingly becoming widely used in citation analysis, thanks to some specific advantages (and despite some disadvantages). Conceptually, through the linkages established, it gives a representation of the basic literature of reference and, incidentally, implies a relation between two papers that reveals a potential common intellectual or methodological approach \citep{Weinberg1974189}. The constancy of the links between the papers over time, being based on cited references which, once published and indexed, is also an asset \citep{Thijs20151453}. Most importantly, the bibliographic coupling is more suitable for recent literature than other citation-based approaches. For reasons of timing and extension of the time window,\footnote{\ Our sampled literature starts in 2003 and ends at the moment the research has been done (November 2020), testifying the recent interest of economic literature on the topic.} using any other citation-based approach would have resulted in a very sparse matrix and created many isolated observations which would not be inter-connected for reasons other than conceptual, but just for the fact that they could not have been cited yet. Not only do the characteristics of our sample motivate the choice of the approach: keeping in mind that this stage of the analysis aims to investigate and map current research fronts in the target literature rather than to look at historical links or the evolution of school of thoughts, bibliographic coupling seems to be the best tool to capture them \citep{Klavans2017984}.

To obtain the network of bibliographically coupled papers, we initially extract the list of cited references from each article and build a bipartite network, a rectangular binary matrix $\mathbf{A}$ linking each paper in the sample to their reference \citep{Aria2017959}:
\begin{equation}
  \mathbf{A} = Paper \times References  
\end{equation}

The matrix $\mathbf{A}$ is composed of 151 rows $i$ representing the papers belonging to the sample and 5.433 columns $j$ representing the ensemble of references cited in each paper in the sample. Each element $a_{ij}$ of the matrix equals 1 when paper $i$ cites paper $j$ in its bibliography; $a_{ij}$ is equal to 0 otherwise. Starting from matrix $\mathbf{A}$, we can derive the bibliographic coupling network $\mathbf{B}$ as follows:
\begin{equation}
\mathbf{B} = \mathbf{A} \times \mathbf{A^T}
\end{equation}
where $\mathbf{A}$ is the cited reference bipartite network and $\mathbf{A^T}$ is its transpose. $\mathbf{B}$ is a symmetrical square matrix 151 $\times$ 151, where rows and columns are papers included in the sample. Element $b_{ij}$ of the matrix $\mathbf{B}$ contains the number of cited articles that paper $i$ and paper $j$ have in common. By construction, the main diagonal will contain the number of references included in each paper (as element $a_{ii}$ defines the number of references that a paper has in common with itself).

The resulting matrix displays an undirected weighted network in which the 151 vertices are the set of papers included in our sample and the edges represent the citation ties between them. An existing tie implies that common reference literature exists between vertex $i$ and $j$. When two nodes are not linked, the corresponding value of their tie is zero, as they do not share any common reference. Therefore, the network is weighted with the strength of the connections between papers $i$ and $j$ being measured by the weights associated with each tie. To avoid loops, which would be meaningless for our investigation,\footnote{\ It is trivial to observe the value of ties that link a paper with itself, which naturally corresponds to the number of listed references.} we set the main diagonal to zero. Few ties exceed 20 shared cited references, with a maximum value of 48.\footnote{\ This number seems very high, but at a closer look, the two papers that register the highest value are two consecutive papers published by the same author \citep{Naude2008, Naude2010}.} It can be argued that the number of references included in an article is not neutral to the resulting tie with any other article. Measuring the correct relatedness of nodes is of primary importance to produce an accurate mapping of literature \citep{Klavans2006251}. Citation behaviors of authors may interfere with the observation of core reference literature at the basis of coupled nodes. An author may opt for an extensive approach of citations and include a consistent number of references to display some particular links or details of a paper; authors may also decide for a less inclusive approach and include just essential cited references in the list. In other words, the number of included references in one article may dissolve meaningful information about the ties. Furthermore, specific formats or types of articles lead to broader or narrower bibliographies.
To address these concerns, a process of normalization is needed so that data can be corrected for differences in the total number of references. Bibliometric literature has dealt with this issue through the calculation of different \textit{similarity measures}. An accurate overview of the possible measures of similarity is provided in \cite{EckNeesJanvan2009Htnc}. Overall, such indices aim to determine the similarity between two units according to their co-occurrence (value of association between them, which in our case, is the number of common references in the bibliography) adjusted in different ways for the number of total occurrences of the single units. However, despite the need to correct data for many purposes in citation-based networks and obtain a size-independent measure of association, there is no consensus on which measure is the most appropriate \citep{EckNeesJanvan2009Htnc}: tests of accuracy and coverage proposed by different authors have reached different conclusions \citep{Klavans2006251,EckNeesJanvan2009Htnc,SternitzkeChristian2009Smfd}. We apply a simple ratio between the observed number of commonly shared references and the product of the number of cited references in each of the two coupled papers. It has been defined as a measure of \textit{association strength} \citep{EckNeesJanvan2009Htnc} and it can be expressed as:

\begin{equation}
 b_{ij}^n = \frac{b_{ij}}{b_{ii}b_{jj}} 
\end{equation}

where $b_{ij}$ corresponds to the weights of the tie between $i$ and $j$ in the original bibliographic coupling network; $b_{ii}$ and $b_{jj}$ are respectively the number of cited references included in paper $i$'s bibliography and in paper $j$'s bibliography, which corresponds to the original value on the diagonal. The obtained weighted network will serve to detect communities of papers through their common references and investigate if referring to a certain (group of) paper(s) creates meaningful clusters of items aggregating around certain common characteristics. 
 
\subsection{Community Detection}

We intend to identify the existence of communities in our network. The assumption is that papers citing the same references aggregate into a group that shares certain features, which could be methodological approach, level of analysis, specific sub-topics of the literature, and outcomes. The extreme heterogeneity of outcomes in this specific literature may be motivated partially by the heterogeneity of the events themselves (type of environmental factor, type of mobility, preexisting conditions in the specific area) or the theoretical and empirical modeling; it may also be motivated by other factors, that can be traced in some patterns linked to the characteristics of single publications. The procedure of community detection is aimed at investigating which are the ``forces" that aggregate or disperse papers with each other, primarily through the direct observation of main characteristics, and then running separate MAs on each cluster. Community detection in the bibliographic network is often made through Louvain community detection algorithm \citep{Blondel2008}. 
In this analysis, a community is thought of as a group of contributions that share common references and form strong common ties with each other, while others have less shared characteristics and structure. The algorithm can detect clusters of contributions with dense interaction with each other and sparse connections with the rest of the network.
\begin{figure}[t]
    \caption{Bibliographic Coupling Network and detected communities}
    \label{fig:network}
    \begin{center}
    \includegraphics[scale = 0.5]{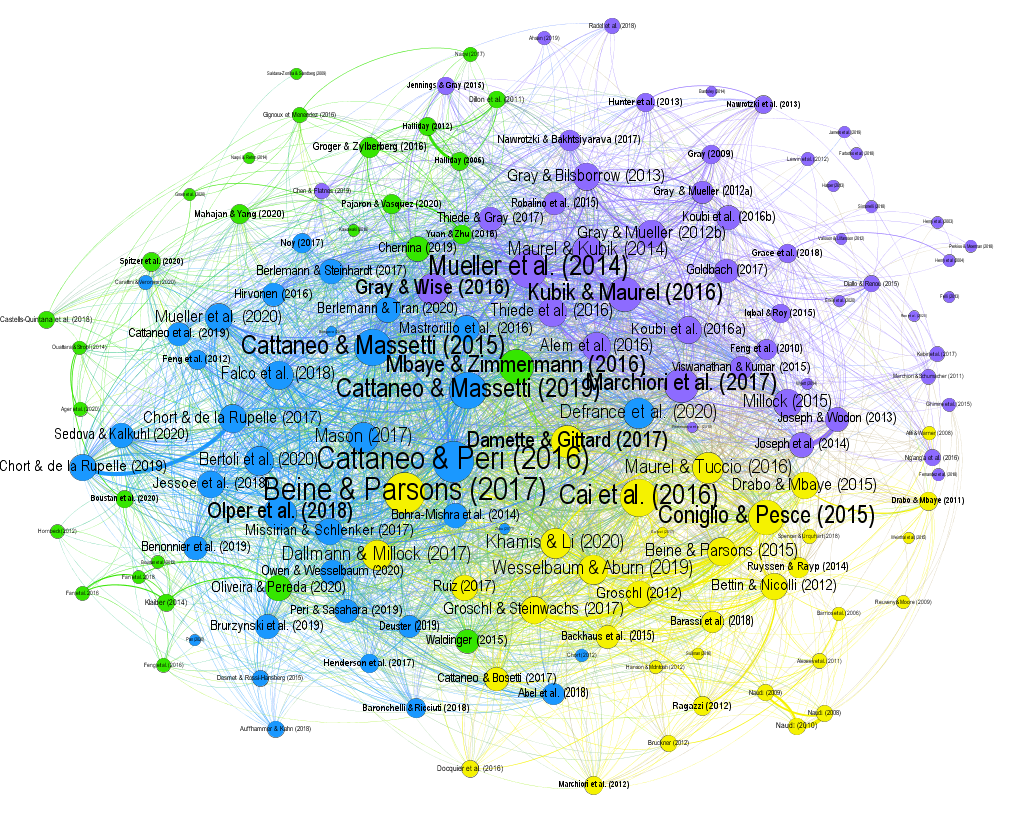}
    \end{center}
    {\scriptsize {\bf Note}: {\linespread{0.5}\selectfont Bibliographic coupling network of 151 documents included in the sample obtained from Scopus, Web of Science, Google Scholar, IDEAS RePEc and previous meta-analyses \citep{hoffmann2020meta, beine_jeusette_2021}. Each node represents a paper included in our sample and its size corresponds to its weighted degree. Nodes are tied by links whenever two nodes share at least one common reference. The thickness of links is given by the association strength of the tie between two nodes (to provide a clear visualization, only nodes with weights higher than the mean are displayed). Colors correspond to communities of belonging of each paper: Cluster 1 is represented in violet, Cluster 2 in green, Cluster 3 in blue, and Cluster 4 in yellow. The description of each Cluster is presented in the text.  \par}}
\end{figure}
 
The procedure identifies four main clusters. Our network being relatively small allows analyzing the main characteristics of each cluster. Following the full-text screening made in the first step of our threefold approach, we summarized some meaningful indicators about the analysis (such as type - quantitative, qualitative, theoretical, policy, literature review -, level - macro or micro for quantitative and qualitative studies -, unit - country, household, individual, territorial units), the object of the analysis (concerning the type of migration and environmental factors studied and the area) and theoretical and empirical approach (empirical approach and whether it is theory-based, estimation strategy and potential channel investigated). Finally, we recorded a synthetic indicator of the concluding effect of environmental factors on migration patterns: for each paper, we assigned the value ``positive", ``negative", ``not significant" or a combination of the three (in case a paper contains multiple analysis of different migration or environmental factors that lead to different outcomes). Thanks to these indicators we were able to have a picture of the main common characteristics of the papers included in a cluster (Appendix Table 4), which will be tested and eventually confirmed in the MA.

The first cluster (Cluster 1) is the most populated, counting 51 papers spanning the entire period considered (from 2003 to 2020). In terms of the type of analysis, it contains the largest variety: as in all clusters, quantitative studies represent the majority (as they are the 76\% of the full sample), but this cluster contains also most of the qualitative analyses (10 out of 13) and policy papers (5 out of 7) of the full sample. Published papers are predominant (47 out of 50). Except for a few papers, the analysis is mainly carried from a micro perspective, with individuals as units of analysis, based on surveys. Interestingly, most of the micro-level studies included in \cite{beine_jeusette_2021} can be found in this cluster. Authorship is very concentrated around two main authors, Clark Gray, (co-)authoring 9 papers, and Valerie Mueller, (co-)authoring 4 papers. Many of their co-authors appear in this community, which indeed scores the highest collaboration index of all communities (2.86), much higher than the full sample (2.16). Another important feature is that Cluster 1 includes the micro-level papers with the highest global citations: \cite{Gray2012pnas}, \cite{Feng2010}, \cite{Gray2012worlddev}, \cite{Mueller2014}, \cite{Henry2004}, \cite{Henry2003} and \cite{Gray2009}. This is also shown by the fact that the number of average citations per document is the highest among all clusters (34.84). Journals are also quite concentrated around a few of them, World Development and Population and Environment mainly. The content of the analyses is mainly focused on climatic change exclusively (precipitation and temperature), while few studies include also natural disasters. All corridors of migration are investigated, with no specific predominance of internal or international migration (which is a characteristic of individual-level studies, mainly based on surveys). Even though the majority of outcomes show a positive coefficient, that can be translated into finding an active role of environmental factors in pushing migrants out of their origin areas, it is not consensual to every paper: variation among results is high compared to other clusters, most paper finding complex relations between the two phenomena and different directions according to different dimensions. Empirical strategies are often based on discrete-time event history models estimated through multinomial logit. This reflects the approach of the main authors included in this community. A strong accent is put on the importance of the agricultural channel and the theme of adaptation to the change in environmental conditions.

The second community (Cluster 2) counts 28 papers, mostly published, except for 4 of them. It is composed of mostly quantitative papers, accompanied by 5 literature reviews. As in the previous cluster, most studies are at a micro level, with all kinds of units of analysis and aggregations. Both patterns of migration are explored, but with special attention to urbanization and internal mobility. Contrarily, it seems to put a stronger accent on natural disasters rather than on slow-onset events. The majority of papers in Cluster 2 have been excluded from \cite{beine_jeusette_2021} (only 5 included, compared to the 21 in Cluster 1) and \cite{hoffmann2020meta} (only 1, all others being in Cluster 4). All papers analyzing the impact of different kinds of natural disasters in the U.S. are included in this cluster. Empirical approaches such as the differences-in-difference model and instrumental variable are often used. The papers explore a large variety of potential channels and mechanisms of transmission of the impact of environmental factors on migration (income, agriculture, employment, liquidity constraints), and only in a few cases, a negative direction is found. 

The third cluster (Cluster 3) includes the most recent papers: only one paper dates 2011, all other ones are published or issued after 2015. This is part of the reasons why the average citations per document in this cluster is the lowest (10.89) compared to any other cluster. Half of the overall unpublished papers are included in this cluster. In terms of kind of analysis, this cluster appears to be very heterogeneous: even if the micro-level analysis is the majority, 12 papers apply a macro-level analysis on countries. Both cross-country and internal migration are considered, but the majority of them investigate the impact of slow-onset events rather than fast-onset. Many of the analyses are theory-based, especially on classic economic migration theories (Roy-Borjas model, New Economics of Labor Migration), or general or partial equilibrium models. This cluster is also peculiar for the heterogeneity of empirical outcomes, which are often multiple for a single paper: outcomes vary according to the different channels explored, i.e. different levels of agricultural dependency, presence of international aid, and level of income. In many cases, environmental factors are an obstacle to the decision to migrate from an area, or completely neutral. Comparatively, outcomes from this cluster tend to show a complex picture and highlight the many dimensions that may intervene in determining the direction of the impact.

Contrary to the previous one, Cluster 4 is extremely homogeneous. It contains almost exclusively quantitative (32 out of 35) and macro-level studies (30 out of 35). It covers equally slow- and fast-onset events and their impact on mobility. Most importantly, it aggregates 23 of the 30 papers reviewed in \cite{hoffmann2020meta}, making this cluster very representative and comparable to \cite{hoffmann2020meta}'s MA. Additionally, this community appears to be solid also in terms of theoretical and empirical approaches, as micro-founded gravity or pseudo-gravity models are widely used in it (more than half of them use such models). None of the studies find a negative impact of environmental factors on migration, they mainly estimate positive and significant outcomes, with few not-significant results for specific cases. The most locally cited macro papers are included in this cluster, which also receive high global citations with an average of citations per document 24.91 (even though lower than Cluster 1). 

This description of cluster composition serves as a preliminary investigation of which are the main characteristics linking papers together through their citation behavior. It emerges that stronger links are given by diverse indicators varying across clusters. To test which are the sources of heterogeneity between clusters that aggregate papers within a cluster and their impact on the estimated effect size, in the next section, we will use this partitioning to run four separate MAs and compare the conclusions.

\section{Meta-Analysis}
\label{sec: meta}

The purpose of our MA is to summarize the results of collected studies and, at the same time, highlight any possible sources of heterogeneity. The analysis is based on four assumptions: (i) our parameter of interest, which we call $\beta$, is the effect of climate change on migration; (ii) most researchers believe that $\beta$ is greater than zero, and this is indeed true; (iii) the sign is not enough for decision-makers; (iv) this has attracted a large literature that has obtained a large number of estimates $\hat{b}$ of $\beta$. Each of the 96 selected papers contains one or more equations that estimate the migration effect due to environmental factors.\footnote{\ Detailed information on collected coefficients and standard errors are provided in the Supplementary material, Section B.} In addition to the characteristics specific to migration itself, the estimated impact on migration can also be distinguished according to different features of environmental factors. Since comparability among studies, and more specifically among estimated $\beta$s, is a crucial issue for the MA, we group all collected estimates and conduct two separate analyses according to the type of environmental phenomenon: gradual or slow-onset events and sudden or fast-onset events. To compare the estimates and correctly interpret the synthetic results we need to standardize all collected effect sizes $\beta$ in a common metric. In this MA the estimates from separate, but similar studies, are converted into partial correlation coefficients ($pcc$):

\begin{equation}
    pcc_i=\frac{t_i}{\sqrt{t_i^2+df_i}},
\end{equation}
and its standard error, $se_i$:

\begin{equation}
se_i=\sqrt{\frac{(1-pcc_i^2)}{df_i}}
\end{equation}

where $t_i$ and $df_i$ are the t-value and the degrees of freedom of the i-th estimate $\beta_i$. 
The $pcc$ is commonly used in MA literature \citep{doucouliagos2005publication,stanley2012meta,  doucouliagos2006economic, brada2021does} and allows to analyze within a single framework of all available studies on the effects of environmental stressors on migration regardless of the specification or measure of migration used.\footnote{\ A summary of the distribution of computed partial correlation coefficients is provided in the Supplementary material, Section C.}
Summarizing all the different estimates together in a single coefficient raises the question of heterogeneity within the same study and between studies. The summary effect is calculated as follows:

\begin{equation}
\hat{\beta}=\frac{\sum_i^N\hat{b}_iw_i}{\sum_i^N w_i},
\end{equation}
where $\hat{b}_i$ is the individual estimate of the effect and weight, $w_i$, in a fixed effects model (FEM) is inversely proportional to the square of the standard error, so that studies with smaller standard errors have greater weight than studies with larger standard errors. The FEM is based on the assumption that the collected effect sizes are homogeneous (the differences observed among the studies are likely due to chance). Unlike in the FEM, random-effects model (REM) takes into account the heterogeneity among studies and weights incorporate a ``between-study heterogeneity", $\hat{\tau}^2$. In the presence of heterogeneity, the two models likely find very different results, and it may not be appropriate to combine results. A test of homogeneity of the $\beta_i$ is provided by referring to the statistic $Q$ to a $\chi^2$-distribution with n - 1 degrees of freedom \citep{higgins2002quantifying}: if the test is higher than the degrees of freedom, the null hypothesis is rejected (and thus there is heterogeneity). Another test commonly used is the $I^2$ inconsistency index by \cite{higgins2002quantifying} describing the percentage of the variability of the estimated effect that is referable to heterogeneity rather than to chance (sample variability). Values of the $I^2$ range from 0 percent to 100 percent where zero indicates no observed heterogeneity. Since most computer programs report $I^2$, and so it is readily available, it is largely used to quantify the amount of dispersion. However, it is a proportion and not an absolute measure of heterogeneity in a meta-analysis \citep{Borenstein2017}. To understand how much the effects vary and report the absolute values, we compute the prediction interval as suggested by \cite{Borenstein2017}. 
The results of the meta-synthesis of the collected estimates (Table \ref{tab:metasynt}) are statistically significant, except for findings of the slow onset effect of paper included in Cluster 2 (where the most of studies focus on the fast onset effect), in which both FEM and REM give statistically insignificant averages.

\begin{table}[t]
\scriptsize
\caption{Basic meta-analysis (Fixed and Random effect MA)}
\label{tab:metasynt}
\begin{tabularx}{\textwidth}{Xclccccc}
\toprule
        &\multicolumn{1}{c}{(1)} &\multicolumn{1}{c}{(2)} &\multicolumn{1}{c}{(3)} &\multicolumn{1}{c}{(4)}  &\multicolumn{1}{c}{(5)} &\multicolumn{1}{c}{(6)} 
        &\multicolumn{1}{c}{(7)} \\
        &      Model	& Averages	&    \makecell[c]{LL prediction \\ interval} &    \makecell[c]{UL prediction \\ interval}	&  I$^2$	 &  \makecell[c]{Q-test\\ p-value}  &  \makecell[c]{N. of Obs.\\ N. of studies}  \\
\midrule
\textit{Slow onset effect} &	FEM	& 0.0001$^{***}$	& -0.0021 &	0.0023   &  &   & 3,897 \\
    & REM	    & 0.0006	    & -0.1000	&	0.1012
    & 99.93 & 0.00 &  66\\
\makecell[c]{- Cluster 1}	& FEM	&0.0001$^{***}$	& -0.0010	&	0.0011	& 	&  & 932\\
	& REM	& -0.0019$^{**}$	& -0.1103	&	0.1065	& 99.98	& 0.00 & 23\\
\makecell[c]{- Cluster 2}	& FEM	& 0.0005	& -0.0042	& 0.0051	& 	&  & 100\\
	& REM	& 0.0152	& -0.1431	&	0.1735	& 99.90	& 0.00 & 3 \\
\makecell[c]{- Cluster 3}	& FEM	& -0.0028$^{***}$	& -0.0250	&	0.0193	& 	&  & 1,814\\
	& REM	& -0.0036$^{***}$	& -0.1236	&	0.1165	& 94.943	& 0.00 & 18\\
\makecell[c]{- Cluster 4}	& FEM	& 0.0049$^{***}$	& -0.0066	&	0.0164 	& 	& & 1,051\\
	& REM	& 0.0066$^{***}$	& -0.0518	&	0.0649	& 95.01	& 0.00  &  22\\
\midrule
\textit{Fast onset effect} &	FEM	& 0.0021$^{***}$	& -0.0125	& 0.0167	& 	&  & 2,032\\
	& REM	& 0.0085$^{***}$	&-0.0964	&	0.1133	& 97.76	& 0.00 & 60\\
\makecell[c]{- Cluster 1}	& FEM	& 0.0015$^{***}$	& -0.0116	&	0.0147	& 	&  & 176\\
	& REM	& 0.0137$^{***}$	&-0.1212	&	0.1485
& 98.87	& 0.00 & 13\\
\makecell[c]{- Cluster 2}	& FEM	& -0.0041$^{***}$	& -0.0191	&	0.0109	& 	&  & 789\\
	& REM	& -0.0054		& -0.0329	&	0.0221	& 57.76	& 0.00 & 16\\
\makecell[c]{- Cluster 3}	& FEM	& -0.0027		& -0.0146	&	0.0200	& 	&  & 409\\
	& REM	& 0.0103$^{***}$	& -0.1274	&	0.1481	& 98.01	& 0.00 & 7 \\
\makecell[c]{- Cluster 4}	& FEM	& 0.0078$^{***}$	& -0.0056	&	0.0213	& 	&  & 688\\
	& REM	& 0.0235$^{***}$	& -0.1070	&	0.1540	& 98.84	& 0.00 & 24 \\
\bottomrule
\multicolumn{8}{@{}m{\textwidth}@{}}{{\bf Note}: {\linespread{0.5}\selectfont Basic meta-analysis of collected estimates. Fixed Effect Model and Random Effect Model are reported for overall slow- and fast-onset samples and sub-samples defined by clusters. Clusters are described in Section 3.2. Averages (2), lower (3) and upper (4) bound of 95\% prediction interval. I$^2$ and Q-test for heterogeneity reported in Columns (4-5); $^{*}$ \(p<0.10\), $^{**}$ \(p<0.05\), $^{***}$ \(p<0.01.\)}}

\end{tabularx}
\end{table}

The preliminary result of the basic MA is that environmental factors seem to influence migration positively, even if the magnitude is very small and the REM mean is statistically significant only in the case of fast-onset events. The mean effect by cluster becomes negative in the case of estimates of slow-onset events in Clusters 1 and 3 and for the estimates of fast-onset events in Cluster 2. 
 
\subsection{Meta-Regression tests of publication selection bias}

Different findings of the same phenomenon can be explained in terms of heterogeneity of studies' features, however, the literature also tends to follow the direction consistent with the theoretical predictions causing the so-called publication bias.\footnote{\ The publication bias occurs when (i) researchers, referees, or editors prefer statistically significant results and (ii) it is easier to publish results that are consistent with a given theory. However, the consequences of the peer-review process refer more to a general ``publication impact" rather than a ``bias" \citep{cipollina2010reciprocal}.} Meta-regression tests, such as the funnel asymmetry test (FAT), allow for an objective assessment of publication bias:

\begin{equation}
\label{eq:simple}
pcc_i=\beta_0 + \beta_1 se_i + \epsilon_i
\end{equation}

Weighted least squares (WLS) corrects the previous equation for heteroskedasticity \citep{stanley2017neither} and it can be obtained by dividing $pcc_i$ by the standard errors:

\begin{equation}
\label{eq:wls}
t_i=\frac{pcc_i}{se_i}=\beta_1 + \beta_0 \frac{1}{se_i} + u_i
\end{equation}

Results are used to test for the presence of publication selection ($H_0:\beta_1  = 0$) or a genuine effect beyond publication selection bias ($H_0:\beta_0 = 0$). According to the Funnel Asymmetry and Precision-Effect Tests (FAT-PET), in the absence of publication selection the magnitude of the reported effect will vary randomly around the “true” value, $\beta_1$, independently of its standard error \citep{stanley2012meta}. Replacing in eq.(\ref{eq:simple}) the standard error $se_i$ with the variance $se_i^2$, as the precision of the estimate, gives a better estimate of the size of the genuine effect corrected for publication bias \citep{stanley2014meta}. This model is called “precision-effect estimate with standard error” (PEESE) and the WLS version is:

\begin{equation}
\label{eq:peese}
t^{PEESE}_i=\frac{pcc_i}{se^2_i}= \beta_1 se_i+\beta_0\frac{1}{se_i}+v_i
\end{equation}

Table \ref{tab:fat_pet} shows results of the FAT-PET using multiple methods for sensitivity analysis and to ensure the robustness of findings. To take into account the issue of the dependence of study results, when multiple estimates are collected in the same study, the errors of meta-regressions are corrected with the ``robust with cluster" option, which adjusts the standard errors for intra-study correlation. 

Column (1) of table \ref{tab:fat_pet} presents the FAT-PET coefficients, column (2) shows the results of the WLS model to deal with heteroskedasticity, columns (3) and (4) present the results of the panel-random effect model (REM) and multilevel mixed-effect model that treats the dataset as a panel or a multilevel structure. 

{\scriptsize {\singlespacing
\setlength\LTleft{0pt}
\setlength\LTright{0pt}
\begin{longtable}{ p{2.4cm} p{4.3cm} cccc}
\caption{FAT-PET results}\\
\toprule\endfirsthead\midrule\endhead\midrule\endfoot\endlastfoot
 & &\makecell[c]{(1)}  &\makecell[c]{(2)} & \makecell[c]{(3)} & \makecell[c]{(4)} \\
 &  & \makecell[c]{WLS} & \makecell[c]{REM} & \makecell[c]{Multilevel\\ Mixed Effect} & \makecell[c]{N. of Obs.}  \\
\midrule
\textit{Slow-onset events}  &   Standard Error (FAT):  $\hat{\beta_1}$  &  	0.108	&  0.268	&  0.260 & \multirow{4}{*}{3,897}\\
&  &  (0.144)	&(0.204)	&(0.208) & \\
	&   Constant (PET):  $\hat{\beta_0}$ 	 &   	0.000$^{*}$	& -0.000	&  -0.000 & \\
&	& (0.000)	& (0.000)	& (0.000) & \\
\midrule
\makecell[c]{- Cluster 1}	&   Standard Error (FAT):  $\hat{\beta_1}$  &  	-0.337	&-0.208	 &-0.213 & \multirow{4}{*}{932}\\
& &	(0.248)	& (0.417)	& (0.407) & \\
&   Constant (PET):  $\hat{\beta_0}$ 	 &	0.000$^{**}$	& 0.000	& 0.000 & \\
& & (0.000)	&(0.000)	&(0.000) & \\
\midrule
\makecell[c]{- Cluster 2}	&   Standard Error (FAT):  $\hat{\beta_1}$  &  	0.412	&0.042	&0.123 & \multirow{4}{*}{100}\\
& & (0.446)	&(0.482)	&(0.488) & \\
&   Constant (PET):  $\hat{\beta_0}$ 	 &	-0.000	&0.000	&-0.000 &\\
& &(0.000)	&(0.000)	&(0.001) & \\
\midrule
\makecell[c]{- Cluster 3}	&   Standard Error (FAT):  $\hat{\beta_1}$  & 	0.001	&0.825*	&0.797** & \multirow{4}{*}{1,814}\\
& & (0.117)	&(0.469)	&(0.357) & \\
&   Constant (PET):  $\hat{\beta_0}$ 	 &-0.004	&-0.011$^{**}$	&-0.011$^{***}$& \\
& & (0.003)	&(0.005)	&(0.001) & \\
\midrule
\makecell[c]{- Cluster 4}	&   Standard Error (FAT):  $\hat{\beta_1}$  & 	0.439	& 0.461 &	0.460 & \multirow{4}{*}{1,051}\\
& & (0.379)	&(0.347)	&(0.443) & \\
&   Constant (PET):  $\hat{\beta_0}$ 	 &0.004$^{**}$	&0.005$^{**}$	&0.005$^{***}$ & \\
& & (0.001)	&(0.002)	&(0.002) & \\
\midrule
\midrule
\textit{Fast-onset events}  &   Standard Error (FAT):  $\hat{\beta_1}$  &  0.532$^{*}$	&0.755$^{**}$	&0.755$^{**}$ &\multirow{4}{*}{2,062}\\
& & (0.274)	&(0.334)	&(0.309) & \\
&   Constant (PET):  $\hat{\beta_0}$ 	 &-0.001	&0.001	&0.001 & \\
& & (0.002)	&(0.002)	&(0.001) & \\
\midrule 							
\makecell[c]{- Cluster 1}	&   Standard Error (FAT):  $\hat{\beta_1}$  & 	0.942$^{**}$	&1.314$^{**}$	&1.329$^{**}$ & \multirow{4}{*}{176}\\
& &(0.366)	&(0.618)	&(0.670) & \\
&   Constant (PET):  $\hat{\beta_0}$ 	 &-0.002	&-0.001	&-0.001& \\
& &(0.002)	&(0.001)	&(0.003) & \\
\midrule								
\makecell[c]{- Cluster 2}	&   Standard Error (FAT):  $\hat{\beta_1}$  & -0.381	&0.095	&0.151 & \multirow{4}{*}{789}\\
& & (0.332)	&(0.410)	&(0.431) &\\
&   Constant (PET):  $\hat{\beta_0}$ 	 &-0.000	&0.001	&0.001 & \\
& & (0.002)	&(0.001)	&(0.002) & \\
\midrule								
\makecell[c]{- Cluster 3}	&   Standard Error (FAT):  $\hat{\beta_1}$  &  	0.283	&0.293	&0.294 & \multirow{4}{*}{409}\\
& &(0.394)	&(0.715)	&(0.372) &\\
&   Constant (PET):  $\hat{\beta_0}$ 	 &-0.002	&0.001	&0.001 & \\
& &(0.004)	&(0.007)	&(0.002)& \\
\midrule								
\makecell[c]{- Cluster 4}	&   Standard Error (FAT):  $\hat{\beta_1}$  & 	1.877$^{**}$	&1.134$^{**}$	&1.072 & \multirow{6}{*}{688}\\
& & (0.703)	&(0.480)	&(0.774) & \\
&   Constant (PET):  $\hat{\beta_0}$ 	 &-0.003	&0.003	&0.003 & \\
& & (0.004)	&(0.005)	&(0.004) & \\
\bottomrule
\multicolumn{6}{@{}m{\textwidth}@{}}{{\bf Note}: {\linespread{0.5}\selectfont FAT, PET coefficients estimated with Weighted Least Squares (1), Random Effect Model (2) and Multilevel mixed effect model. Overall effect of slow- and fast-onset events reported, along with sub-samples defined by clusters. PCC precision square weights ($1/se_i^2 $); robust standard errors clustered by study in parentheses; $^{*}$ \(p<0.10\), $^{**}$ \(p<0.05\), $^{***}$ \(p<0.01.\)}}\\
\label{tab:fat_pet}
\end{longtable}
}}

Looking at the estimates of the effect of climate change on migration, the FAT coefficients ($\hat{\beta}_1$) are not statistically significant, implying that there is no evidence of publication bias, while the positive and statistically significant PET coefficient ($\hat{\beta}_0$) indicates a genuinely positive slow-onset effect exists, in particular in the case of Cluster 4. Conversely, in the case of Cluster 3 the REM and multilevel mixed-effect model find that, even if in presence of publication bias, the impact on migration is negative. Table \ref{tab:fat_pet} provides evidence of publication bias in the literature focusing on the effect of natural disasters on migration. The estimated FAT coefficient is statistically significant in the overall sample, especially due to papers in clusters 1 and 3, and there is insufficient evidence of a genuinely positive effect (accept $H_0: \hat{\beta}_0$).

\subsection{Multiple Meta-Regression Analysis: econometric results and discussion}

The multiple meta-regression analysis (MRA) includes an encompassing set of controls for factors that can integrate and explain the diverse findings in the literature. To capture possible sources of bias among all analyses, we code all differences in the features of the various studies and regressions and include a set of dummies to control for them.  Specifically, we code left- and right-hand side characteristics of regressions estimated in the collected papers and generate a set of dummies for paper features, dependent variables, independent variables, sample characteristics, and regression characteristics.\footnote{\ The complete description of coded variables is available in Supplementary material, section D.}

The overall sample includes both unpublished and published papers, so we add some moderators variables describing different features of the studies that are published. In particular, we introduce a dummy for \textit{Published articles} and a control for the quality of the journal in which the study is published by adding the variable \textit{Publication Impact-factor}. In reporting the main results, some authors emphasize a benchmark regression that produces a preferred estimate, thus we add the dummy \textit{preferred specification} equal to 1 when the reported effect size is obtained from the main specification. Concerning the measure of migration, the dependent variable in the left-hand side of the regression, original studies mainly distinguish migration by \textit{corridor}, which are mainly two, internal and international migration. In this context, we distinguish also a special internal corridor, the one characterized by rural-urban mobility, to investigate the potential impact of an environmental variable on the urbanization process. Whenever the corridor is not specified, the variable is categorized as undefined (which will be the reference category in the estimation). Dependent variables differ also in terms of \textit{measurement} of the phenomenon: specifically, we separate measures that express flows from those expressing stocks. The first category includes both studies that use flows (or an estimation of flows) and rates of migration. The second category captures those cases in which migration is measured as a stock of migrants at the destination. The reference category is direct measures, which mainly capture whether migration has occurred or not (typically dummy variables used on survey-based samples equal to 1 when the individual migrates and 0 otherwise). We also include information about the countries of origin and the destination of migrants. \textit{Origins} are categorized by macro-regions: Africa, Asia, Europe, Latin America and Caribbean, Middle East and North Africa, and North America. The reference category is ``world", identified whenever origin countries are not specified (typically in multi-country settings). \textit{Destinations} are categorized by level of income. The choice of this categorization is led by the aim to identify differences in the possibility to choose a destination. Categories are divided into high, higher-middle, lower-middle, and low-income. 

The specific objective of the study is the impact of environmental variables on migration, thus on the right-hand side of the regression a proxy of the environmental change is included. Slow-onset events are typically defined as gradual modifications of temperature, precipitation, and soil quality. Respectively, three dummies \textit{temperature, precipitation} and \textit{soil degradation} are created. Each of these phenomena is measured in different ways, and the use of a specific kind of measurement is relevant to the outcome. Both temperature and precipitation have been measured in levels (simple level or trend of temperature/precipitation); deviation, as the difference between levels and long-run averages; and anomalies, mostly calculated as the ratio of the difference between the level and the long-run mean and its standard deviation. Soil degradation includes events such as desertification, soil salinity, or erosion. Additionally, we also code the time lag considered concerning the time units of the dependent variable: whenever the period considered corresponds to the same period of the dependent variable the lag is zero, while it takes values more than zero for any additional period before the dependent variable time-span. This control also allows us to account for varying time-frames in different studies, including situations where migration spans several years or occurs suddenly in the aftermath of a natural disaster.  The second battery of coded variables refers to fast-onset events, which can be also defined as natural hazards or extreme events. The main classification of fast-onset events reflects the one reported in Section \ref{sec: review}: \textit{geophysical} (earthquakes, mass movements, volcanic eruptions), \textit{meteorological} (extreme temperature, storms - cyclones, typhoons, hurricanes, tropical storms, tornadoes), \textit{hydrological} (floods and landslides) and \textit{climatological} (droughts or wildfires). Fast-onset events also differ in the way they are measured. Possible measures are occurrence (when the measure is a dummy capturing if the disaster happened or not), frequency (the count of events that occurred in the area), intensity (i.e. Richter scale for earthquakes, wind speed for tornadoes, etc.), duration (length of the occurrence of the event) and losses (when the disaster is measured in terms of the affected population, number of deaths or injured people, number of destroyed houses or financial value of the damaged goods). As for slow-onset events, we code a continuous variable capturing the time lag of the event concerning the dependent variable. A dummy capturing whether the coefficient refers to multiple disasters is also included.

Characteristics of the sample are one of the main sources of heterogeneity. The level of the analysis varies considerably from paper to paper, as we include both micro-and macro-level studies. we code variables capturing both the specific unit of analysis and the source of the data. Typically micro-level studies use data coming from \textit{censuses} or \textit{surveys} where \textit{households} or \textit{individuals} are the units of analysis. \textit{Country-level} studies usually take the source of their data from \textit{official statistics}. Other kinds of sampling are included in the reference group (for example small territorial aggregates such as districts, provinces, or grid cells). We also code a variable capturing the time span of the analysis, subtracting the last year of observation from the first one. The role of econometric approaches may have an impact on resulting outcomes. \cite{beine_jeusette_2021} emphasized in their work the importance of methodological choices, with differentiated results depending on estimation techniques. First of all, we code a \textit{panel} dummy to capture whether the structure of data and related estimation techniques has an impact. Furthermore, we distinguish \textit{Poisson} estimations that include the Pseudo Poisson Maximum Likelihood (PPML) estimator and Negative Binomial Models; \textit{linear} estimators, both Ordinary Least Squares (OLS), linear probability models and maximum likelihood models; conventional \textit{Instrumental Variables} (IV) estimators, two-stage least squares (2SLS), and other cases of estimators as Generalized Method of Moments (GMM) used to control for endogeneity; and finally, \textit{logit} which comprises multinomial logit models. Any other estimator (i.e. Tobit, panel VAR) is less frequent and grouped in a category \textit{other estimators} used as the reference group. 

Theoretically, the impact of environmental variables on migration may be mediated, channe-led, or transmitted through other phenomena that can be controlled for or interacted with. Most of models investigating general migration determinants usually control for several possible determinants to recover the effect of the specific objective variable, with all potential other factors being controlled for. The majority of these additional controls are suggested by theoretical models and then introduced in the empirical model. Furthermore, methodological approaches in our sample are found to often include interaction terms to specifically address the combined effect of an environmental variable with other potential factors. Thus, we introduce two groups of variables, \textit{controls} and \textit{interacted terms}, categorized both to capture factors or channels such as income, agriculture, conflicts, political stability, cultural or geographical factors. Among the list of controls, we also include a dummy that captures whether both slow- and fast-onset events are included in the regression.

Table \ref{tab:MRA_slow_stepwise} shows the results of the multiple MRA on the literature in slow-onset events (precipitation, temperature, and soil quality) in which potential biases are filtered out sequentially by the addition, in a stepwise manner, of statistically significant controls.
Column (1) presents results for the whole sample of studies estimating the impact of climatic variations on migration, and columns (2) to (5) show the results of papers grouped by clusters to highlight how specific features characterizing the cluster influence the magnitude of the estimated effect. The results are unfolded below.

{\scriptsize {\singlespacing
\begin{longtable}{p{5cm} ccccc}
\caption{MRA Results for slow-onset events}\\
\label{tab:MRA_slow_stepwise}\\
\toprule\endfirsthead\midrule\endhead\midrule\endfoot\endlastfoot
        &\multicolumn{1}{c}{(1)} &\multicolumn{1}{c}{(2)} &\multicolumn{1}{c}{(3)} &\multicolumn{1}{c}{(4)}     &\multicolumn{1}{c}{(5)} \\
        &      All &Cluster 1 &Cluster 2 &Cluster 3 &Cluster 4 \\
\midrule
Constant (PET): $\hat{\beta_0}$&   -0.011$^{***}$&   -1.040$^{***}$&    0.959$^{***}$&   -0.031$^{***}$&    0.102$^{***}$\\
        &  (0.003) &  (0.264) &  (0.010) &  (0.008) &  (0.009) \\
Standard Error (FAT): $\hat{\beta_1}$&   -0.205$^{*}$  &   -4.939$^{**}$ &  -29.959$^{***}$&    0.099 &   -0.671$^{***}$\\
        &  (0.119) &  (1.894) &  (0.264) &  (0.273) &  (0.190) \\
        \midrule
\textit{Paper features}&  &  &  &  &  \\
 - Preferred specification&  &   -0.001$^{**}$ &  &  &  \\
        &  &  (0.000) &  &  &  \\
 - Published article&  &  &  &  &   -0.008$^{***}$\\
        &  &  &  &  &  (0.002) \\
 - Publication Impact-factor&  &    0.024$^{**}$ &  &  &  \\
        &  &  (0.009) &  &  &  \\
        \midrule
\textit{Corridor}&  &  &  &  &  \\
 - Internal     &    0.002$^{***}$&    0.002$^{***}$&  &   -0.009$^{***}$&    0.012$^{**}$ \\
        &  (0.001) &  (0.000) &  &  (0.002) &  (0.005) \\
 - International&  &  &  &   -0.010$^{***}$&  \\
        &  &  &  &  (0.001) &  \\
 - Urbanisation &    0.002$^{***}$&    0.002$^{***}$&  &  &  \\
        &  (0.001) &  (0.000) &  &  &  \\
\textit{Measurement}&  &  &  &  &  \\
 - Flows        &   -0.016$^{***}$&    1.565$^{***}$&  &  &  \\
        &  (0.004) &  (0.481) &  &  &  \\
\textit{Region of origin}&  &  &  &  &  \\
 - Asia &    0.008$^{**}$ &  &  &  &  \\
        &  (0.003) &  &  &  &  \\
 - Europe       &    0.033$^{***}$&   -0.332$^{***}$&  &  &    0.010$^{***}$\\
        &  (0.004) &  (0.083) &  &  &  (0.002) \\
 - LAC  &  &  &  &    0.096$^{***}$&   -0.012$^{***}$\\
        &  &  &  &  (0.017) &  (0.002) \\
 - North America&   -0.021$^{***}$&  &  &  &  \\
        &  (0.004) &  &  &  &  \\
\textit{Destination}&  &  &  &  &  \\
 - High income  &  &    &  &   -0.049$^{***}$&  \\
        &  &   &  &  (0.012) &  \\
 - Upper-middle income&  &   &  &   -0.049$^{***}$&  \\
        &  &   &  &  (0.012) &  \\
 - Lower-middle income&  &    &  &    0.004$^{***}$&  \\
        &  &  &  &  (0.001) &  \\      
        \midrule
\textit{Precipitation measures}&  &  &  &  &  \\
 -  levels      &  &    &   -0.924$^{***}$&   -0.007$^{***}$&   -0.002$^{*}$  \\
        &   &   &  (0.009) &  (0.002) &  (0.001) \\
 -  deviation   &  &    &  &   -0.008$^{**}$ &  \\
        &  &  &  &  (0.004) &  \\
 -  anomaly     &  &    0.002$^{**}$ &  &  &  \\
        &  &  (0.001) &  &  &  \\
\textit{Temperature measures}&  &  &  &  &  \\
 - levels       &  &    &   -0.924$^{***}$&  &  \\
        &  &   &  (0.009) &  &  \\
 - deviation    &   &   &   -0.410$^{***}$&  &  \\
        &   &   &  (0.005) &  &  \\
 - anomaly      &  &   -0.005$^{***}$&  &   -0.012$^{***}$&  \\
        &  &  (0.001) &  &  (0.001) &  \\
Time lag        &  &  &    0.021$^{***}$&  &  \\
        &  &  &  (0.000) &  &  \\
Soil Degradation&  &    0.011$^{***}$&  &   -0.055$^{***}$&  \\
        &  &  (0.003) &  &  (0.002) &  \\
            \midrule
\textit{Sample features}&  &  &  &  &  \\
Time span       &   &  &  &   -0.002$^{***}$&  \\
        &  &  &  &  (0.000) &  \\
\textit{Source of data}&  &  &  &  &  \\
 - Census       &    0.016$^{***}$&   -0.331$^{**}$ &  &    0.076$^{***}$&   -0.089$^{***}$\\
        &  (0.002) &  (0.140) &  &  (0.012) &  (0.005) \\
 - Official statistics&  &    0.397$^{***}$&  &  &  \\
        &  &  (0.096) &  &  &  \\
 - Research data&   -0.007$^{**}$ &    0.257$^{**}$ &  &  &  \\
        &  (0.003) &  (0.103) &  &  &  \\
\textit{Unit of analysis}&  &  &  &  &  \\
 - Household    &  &    1.256$^{***}$&  &    0.052$^{***}$&  \\
        &  &  (0.362) &  &  (0.005) &  \\
 - Individual   &   -0.015$^{***}$&    1.051$^{***}$&  &  &  \\
        &  (0.004) &  (0.287) &  &  &  \\
 - Country level&    0.014$^{***}$&   -0.856$^{**}$ &  &    0.079$^{***}$&   -0.098$^{***}$\\
        &  (0.004) &  (0.311) &  &  (0.019) &  (0.009) \\
        \midrule
\textit{Estimation:}&  &  &  &  &  \\
 - Panel        &    0.019$^{***}$&    0.066$^{**}$ &  &    0.042$^{***}$&  \\
        &  (0.004) &  (0.024) &  &  (0.006) &  \\
 - Poisson      &  &   -0.514$^{**}$ &  &  &  \\
        &  &  (0.219) &  &  &  \\
 - OLS and ML   &    0.010$^{***}$&  &   -0.017$^{***}$&  &    0.011$^{***}$\\
        &  (0.003) &  &  (0.000) &  &  (0.002) \\
 - IV   &    0.041$^{***}$&  &  &  &    0.044$^{***}$\\
        &  (0.011) &  &  &  &  (0.011) \\
        \midrule
\textit{Controls:}&  &  &  &  &  \\
 - Slow and fast included&  &  &  &   -0.032$^{***}$&  \\
        &  &  &  &  (0.008) &  \\
 - Income       &    0.004$^{***}$&    0.170$^{**}$ &    0.004$^{***}$&  &  \\
        &  (0.001) &  (0.065) &  (0.000) &  &  \\
 - Conflict     &  &    0.249$^{***}$&  &  &  \\
        &  &  (0.063) &  &  &  \\
 - Political stability&  &   -0.130$^{***}$&  &  &    0.012$^{***}$\\
        &  &  (0.040) &  &  &  (0.002) \\
 - Population   &    0.005$^{**}$ &  &  &    0.031$^{***}$&    0.009$^{***}$\\
        &  (0.002) &  &  &  (0.008) &  (0.002) \\
 - Diaspora     &  &   -0.156$^{**}$ &  &  &  \\
        &  &  (0.074) &  &  &  \\
 - Past migration&  &   -0.090$^{**}$ &    0.007$^{***}$&  &  \\
        &  &  (0.042) &  (0.000) &  &  \\
 - Poverty      &  &    0.096$^{**}$ &  &  &   -0.011$^{***}$\\
        &  &  (0.039) &  &  &  (0.002) \\
 - Culture      &  &    0.436$^{**}$ &  &  &  \\
        &  &  (0.173) &  &  &  \\
 - Agriculture  &    0.004$^{***}$&   -0.461$^{**}$ &  &  &  \\
        &  (0.001) &  (0.193) &  &  &  \\
 - Labour       &  &  &  &  &  \\
        &  &  &  &  &  \\
 - Urban        &   -0.013$^{***}$&    0.265$^{**}$ &  &   -0.016$^{***}$&  \\
        &  (0.002) &  (0.111) &  &  (0.004) &  \\
 - International aids&   -0.025$^{***}$&  &  &   -0.036$^{***}$&  \\
        &  (0.008) &  &  &  (0.003) &  \\
        \midrule
\textit{Interacted terms (channels):}&  &  &  &  &  \\
 - Agriculture  &  &  &   -0.055$^{***}$&  &    0.003$^{*}$  \\
        &  &  &  (0.000) &  &  (0.001) \\
 - International aid&    0.023$^{*}$  &  &  &    0.034$^{***}$&  \\
        &  (0.013) &  &  &  (0.000) &  \\
 - Culture      &   -0.006$^{***}$&  &  &  &   -0.006$^{***}$\\
        &  (0.001) &  &  &  &  (0.002) \\
 - Destination  &    0.012$^{***}$&  &  &  &  \\
        &  (0.002) &  &  &  &  \\
 - Poverty      &  &  &  &   -0.058$^{***}$&  \\
        &  &  &  &  (0.011) &  \\
 - Income and agriculture&    0.029$^{***}$&  &  &    0.024$^{***}$&  \\
        &  (0.005) &  &  &  (0.004) &  \\
 - Environment  &  &  &  &    0.003$^{*}$  &  \\
        &   &   &  &  (0.001) &  \\
 - Income       &  &   -0.003$^{**}$ &  &   -0.018$^{***}$&  \\
        &  &  (0.001) &  &  (0.004) &  \\
 - Origin       &  &   &  &   -0.046$^{***}$&  \\
        &   & &  &  (0.005) &  \\
 - Past migration&   -0.013$^{***}$&   -0.007$^{***}$&  &  &  \\
        &  (0.003) &  (0.000) &  &  &  \\
 - Political stability&   -0.037$^{***}$&  &  &  &   -0.047$^{***}$\\
        &  (0.008) &  &  &  &  (0.002) \\
 - Population   &   -0.019$^{***}$&  &  &   -0.028$^{***}$&  \\
        &  (0.006) &  &  &  (0.008) &  \\
 - Urban        &    0.011$^{***}$&  &  &    0.021$^{***}$&  \\
        &  (0.004) &  &  &  (0.001\\
\midrule
PEESE Correction: $\beta_0$&   -0.012$^{***}$&          -0.783$^{***}$&           0.655$^{***}$&          -0.030$^{***}$&           0.079$^{***}$\\
          & [-0.018,-0.006]         & [-1.155,-0.411]         &   [0.582,0.729]         & [-0.038,-0.022]         &   [0.062,0.097]         \\
\midrule
N. of Obs.     &     3,897 &      932 &      100 &     1,814 &     1,051 \\
N. of Studies     &     66 &      23 &      3 &      18 &      22 \\\bottomrule
\multicolumn{6}{p{\dimexpr0.93\linewidth+3\tabcolsep}}{
{\bf Note}: {\linespread{0.5}\selectfont Step-wise regression of overall sample (1) and sub-samples defined by clusters (2-5) for slow-onset events. Estimates shown represent significant coefficients obtained through a step-wise procedure (not reported when not significant). Some coefficients and standard errors are zeros because of rounding. Controls are grouped by paper features, dependent variable, independent variable, sample and regression characteristics. PCC precision square weights ($1/se_i^2 $); robust standard errors clustered by study in parentheses; 95$\%$ confidence intervals in brackets; $^{*}$ \(p<0.10\), $^{**}$ \(p<0.05\), $^{***}$ \(p<0.01.\)  }}
\end{longtable}
}}

Column (1) refers to the overall sample and shows a coefficient of the main variable of interest ($\hat{\beta}_0$) negative and statistically significant, implying that climatic variations may decrease incentives for migration by exacerbating credit constraints of potential migrants. Looking at results for different clusters (columns 2-5) such a negative effect is generated by studies that are included in clusters 1 and 3. The MRA of papers in clusters 2 and 4, instead, gives positive and statistically significant PET coefficients  ($\hat{\beta}_0$) implying that climate changes induce people to migrate. Concerning the FAT-test, the intercept ($\hat{\beta}_1$) might deviate from zero confirming the presence of publication bias: the peer-review process seems to particularly affect the magnitude of the estimated effect of studies in all clusters except for Cluster 3. 

Most of the papers included in the MRA for slow-onset events are published (52 articles out of 66), indeed the estimated coefficients of controls for published articles are useful to evaluate if the peer-review process exerts some influence on reported results in the collected studies. In Cluster 3 estimates obtained by the \textit{Preferred specification} tend to be slightly lower while articles published in journals with higher impact factors report lower estimates of the impact of slow-onset events on migration. In Cluster 4, instead, results of \textit{Published articles} are lower, even if the mean effect of this group of studies remains positive.

From the other sets of controls emerges that specific features of studies included in the MRA differently explain the diversity in the results within clusters. The positive coefficients of controls for corridors such as \textit{Internal} and \textit{Urbanization} state that people respond to adverse climatic change with increased internal migration.  The only exception is for studies included in Cluster 3, this is the most heterogeneous cluster of most recent papers, where heterogeneous approaches (micro-and macro-level and type of migration) lead to a large heterogeneity in outcomes, varying according to different channels explored. Findings obtained when mobility is measured by \textit{Flows} seem to be lower in the overall sample. In macroeconomic literature, usually, the measurement of migration is a stock variable, since it is generally easier to find and measure the number of foreign citizens born or resident in a country at any given time. Data on flow variables and migration rates, or the number of people who have moved from an origin to a destination in a specific period, are less available, and analyses often rely on estimates and computations of this data. Therefore, the opposite sign of the coefficient of the variable \textit{Flows} in Cluster 1 is not surprising since this cluster collects all micro-level studies (where the migration variable refers to the movements of individuals as a unit, based on surveys).

Controls for how the climatic phenomenon is measured, \textit{Precipitation measures} and \textit{Tempera-ture measures}, seem to differently affect the heterogeneity of results and, in many cases, the estimated coefficients are statistically significant but very close to zero. 

The estimated coefficients of dummies for country groups included in our multiple MRA indicate how results from analyses focusing on specific regions of origin differ. In particular, positive coefficients of controls \textit{Asia} and \textit{Europe} support the idea that the results of analyses that focus on the migration from these regions are likely to be positive (with exception of Cluster 1), while if the people move from a country in the region of \textit{North America} the impact of climate changes on migration is lower and can be negative. The climate impact on migration from \textit{LAC} (Latin America and the Caribbean) countries are higher in Cluster 3 (where the PET coefficient is negative) and lower in Cluster 4 (where the PET coefficient is positive). 

Regarding the heterogeneity produced by the fact that studies use different sources of data for migration, we add dummies for sources used. All estimated coefficients of this set of controls are statistically significant in Cluster 1: the use of different databases might influence the wide variety of findings. Effect sizes in Cluster 2, instead, are not affected by the source of data used. 

Since it is natural to expect the adjustment of migratory flows in response to climate change is not instantaneous, especially in the case of gradual phenomena, most of the studies use a panel structure with a macroeconomic focus and attempt to assess the impact of changes in climatic conditions on human migratory flows in the medium-long term. Microeconomic analyses mostly use cross-section data to explain causal relationships between specific features of individuals, collected through surveys and censuses, and various factors determining migration by isolating the net effect of the environment. Analyses at \textit{Individual} level tend to capture a more negative impact of climate changes on migration, whereas analyses at \textit{Country} level tend to find a more positive effect. As already said, for micro-level analyses in Cluster 1 controls related to sample characteristics have opposite signs.
Looking at dummies for the estimation techniques, our evidence suggests that the diversity in the effect sizes is in part explained by differences in techniques. In particular, positive and significant coefficients are found for controls as \textit{OLS and ML} estimators for cross-section analyses, same for panel studies that use \textit{Panel} estimation techniques, and Instrumental Variables (\textit{IV}) or GMM estimators to correct for endogeneity. Micro-economic analyses (Cluster 1) use more disaggregated data, while the high presence of zeros in the dependent variable is treated with a \textit{Poisson} estimator, which tends to produce lower estimates. 

Many authors highlight the importance of variables of political, economic, social, and historical nature, in influencing the impact of climatic anomalies on migration processes, emphasi-zing the role of important channels of transmission of the environmental effect to migrations. We include in the multiple MRA a set of dummies for \textit{Controls} included in the estimation of the model of migration and dummies for \textit{Channels} through which the climatic event determines migration phenomena. The idea is that studies based on the same theoretical framework tend to include the same set of control variables or interacted terms and we find that many of these controls may positively and negatively affect the effect size of climate changes on migration.

Table \ref{tab:MRA_fast_stepwise} shows the results of the MRA for fast-onset events, or rather natural disasters, more or less related to climate change, which appear as destructive shocks of limited duration and for which the ability to predict is reduced.\footnote{\ Potential biases are filtered out sequentially by the addition, in a stepwise manner, of statistically significant controls.}

{\scriptsize {\singlespacing
\begin{longtable}{p{5cm} ccccc}
\caption{MRA Results for fast-onset events}\\
\toprule\endfirsthead\midrule\endhead\midrule\endfoot\endlastfoot
        &\multicolumn{1}{c}{(1)} &\multicolumn{1}{c}{(2)} &\multicolumn{1}{c}{(3)} &\multicolumn{1}{c}{(4)} &\multicolumn{1}{c}{(5)} \\
        &      All &Cluster 1 &Cluster 2 &Cluster 3 &Cluster 4 \\
\midrule
Constant (PET): $\hat{\beta_0}$&    0.044$^{**}$ &   -0.127$^{***}$&    3.147$^{***}$&   -0.508$^{***}$&    0.419$^{***}$\\
        &  (0.021) &  (0.032) &  (0.091) &  (0.038) &  (0.030) \\
Standard Error (FAT): $\hat{\beta_1}$&    0.997$^{***}$&   -1.506 &   -0.097 &    6.410$^{***}$&    1.070 \\
        &  (0.279) &  (1.399) &  (0.116) &  (0.961) &  (0.783) \\
        \midrule
\textit{Paper features}&  &  &  &  &  \\
 - Preferred specification&  &  &  &    0.001$^{***}$&  \\
        &  &  &  &  (0.000) &  \\
 - Published articles&  &    0.145$^{***}$&    0.936$^{***}$&  &  \\
        &  &  (0.004) &  (0.056) &  &  \\
 - Publication Impact-factor&    0.002$^{**}$ &    0.015$^{***}$&   -0.475$^{***}$&  &    0.048$^{*}$  \\
        &  (0.001) &  (0.004) &  (0.007) &  &  (0.026) \\
        \midrule
\textit{Corridor}&  &  &  &  &  \\
 - Internal     &  &  &  &    0.043$^{***}$&   -0.021$^{**}$ \\
        &  &  &  &  (0.005) &  (0.008) \\
 - International&  &  &    0.004$^{***}$&    0.041$^{***}$&  \\
        &  &  &  (0.001) &  (0.005) &  \\
 - Urbanization &  &  &    0.003$^{***}$&  &  \\
        &  &  &  (0.000) &  &  \\
\textit{Measurement}&  &  &  &  &  \\
 - Flows        &  &    0.322$^{***}$&   -3.199$^{***}$&   -0.240$^{***}$&   -0.355$^{***}$\\
        &  &  (0.027) &  (0.296) &  (0.042) &  (0.072) \\
 - Stock        &  &  &   -0.087$^{***}$&  &   -0.357$^{***}$\\
        &  &  &  (0.012) &  &  (0.071) \\
\textit{Region of Origin}&  &  &  &  &  \\
 - Africa       &   -0.015$^{**}$ &   -0.003$^{**}$ &    0.346$^{***}$&    0.212$^{***}$&  \\
        &  (0.007) &  (0.001) &  (0.106) &  (0.044) &  \\
 - Asia &  &  &   -0.773$^{***}$&  &  \\
        &  &  &  (0.145) &  &  \\
 - Europe       &  &   -0.340$^{**}$ &    2.114$^{***}$&  &  \\
        &  &  (0.156) &  (0.313) &  &  \\
 - LAC  &  &   -0.034$^{***}$&    0.974$^{**}$ &    0.030$^{***}$&  \\
        &  &  (0.002) &  (0.380) &  (0.001) &  \\
 - North America&   -0.023$^{**}$ &  &    1.827$^{***}$&  &  \\
        &  (0.009) &  &  (0.332) &  &  \\
\textit{Destination}&  &  &  &  &  \\
 - High income  &  &  &   -4.148$^{***}$&  &   -0.003$^{*}$  \\
        &  &  &  (0.181) &  &  (0.002) \\
 - Upper-middle income&  &  &  &   -0.003$^{*}$  &  \\
        &  &  &  &  (0.001) &  \\
 - Lower-middle income&  &   -0.002$^{***}$&   -0.002$^{***}$&    0.021$^{***}$&   -0.020$^{***}$\\
        &  &  (0.000) &  (0.000) &  (0.000) &  (0.004) \\        
        \midrule
\textit{Type of event}&  &  &  &  &  \\
 - Geophysical  &  &  &   -0.054$^{***}$&   -0.107$^{***}$&  \\
        &  &  &  (0.002) &  (0.006) &  \\
 - Meteorological&    0.004$^{**}$ &  &   -0.063$^{***}$&   -0.146$^{***}$&  \\
        &  (0.002) &  &  (0.006) &  (0.006) &  \\
 - Hydrogeological&    0.005$^{**}$ &    0.006$^{**}$ &   -0.054$^{***}$&   -0.109$^{***}$&    0.006$^{**}$ \\
        &  (0.002) &  (0.002) &  (0.002) &  (0.006) &  (0.003) \\
 - Climatological&  &  &   -0.065$^{***}$&   -0.077$^{***}$&  \\
        &  &  &  (0.006) &  (0.006) &  \\
Time lag        &  &  &    0.002$^{***}$&  &  \\
        &  &  &  (0.000) &  &  \\
\textit{Measurement}&  &  &  &  &  \\
 - Frequency    &  &    0.031$^{***}$&   -0.023$^{***}$&    0.556$^{***}$&  \\
        &  &  (0.000) &  (0.000) &  (0.026) &  \\
 - Intensity    &  &  &    1.137$^{***}$&    0.493$^{***}$&  \\
        &  &  &  (0.265) &  (0.026) &  \\
 - Occurrence   &  &  &    0.024$^{***}$&    0.474$^{***}$&  \\
        &  &  &  (0.000) &  (0.009) &  \\
 - Duration     &  &    0.368$^{***}$&  &    0.584$^{***}$&  \\
        &  &  (0.057) &  &  (0.029) &  \\
            \midrule
\textit{Sample} &  &  &  &  &  \\
Time span       &   &    0.014$^{***}$&    0.030$^{***}$&   -0.001$^{*}$  &  \\
        & &  (0.003) &  (0.005) &  (0.000) &  \\
\textit{Source of data}&  &  &  &  &  \\
 - Census       &  &  &   -0.005$^{***}$&  &  \\
        &  &  &  (0.000) &  &  \\
 - Official statistics&  &   -0.127$^{**}$ &    0.002$^{***}$&  &    0.152$^{*}$  \\
        &  &  (0.044) &  (0.000) &  &  (0.085) \\
 - Research data&  &  &  &  &  \\
        &  &  &  &  &  \\
 - Survey       &  &  &   -3.360$^{***}$&  &  \\
        &  &  &  (0.052) &  &  \\
\textit{Unit of analysis}&  &  &  &  &  \\
 - Household    &  &   -0.197$^{***}$&   -0.910$^{***}$&    0.757$^{***}$&  \\
        &  &  (0.064) &  (0.027) &  (0.067) &  \\
 - Individual   &  &  &    0.121$^{***}$&  &  \\
        &  &  &  (0.032) &  &  \\
 - Country level&  &  &  &  &   -0.230$^{*}$  \\
        &  &  &  &  &  (0.116) \\
        \midrule
\textit{Estimation}&  &  &  &  &  \\
 - Panel        &   -0.034$^{***}$&   -0.621$^{***}$&    0.788$^{***}$&  &   -0.116$^{*}$  \\
        &  (0.011) &  (0.103) &  (0.059) &  &  (0.059) \\
 - Poisson      &  &  &  &   -0.003$^{***}$&    0.058$^{***}$\\
        &  &  &  &  (0.000) &  (0.010) \\
 - OLS and ML   &   -0.027$^{**}$ &   -0.037$^{***}$&  &  &    0.036$^{***}$\\
        &  (0.012) &  (0.003) &  &  &  (0.011) \\
 - IV   &   -0.066$^{***}$&   -0.037$^{***}$&    0.830$^{***}$&  &    0.058$^{*}$  \\
        &  (0.019) &  (0.003) &  (0.043) &  &  (0.031) \\
 - Logit        &   -0.023$^{*}$  &  &  &  &  \\
        &  (0.012) &  &  &  &  \\
        \midrule
\textit{Controls}&  &  &  &  &  \\
 - Slow and fast included&   -0.016$^{*}$  &  &  &  &  \\
        &  (0.009) &  &  &  &  \\
 - Income       &  &  &    0.008$^{***}$&   -0.009$^{***}$&    0.094$^{*}$  \\
        &  &  &  (0.000) &  (0.000) &  (0.049) \\
 - Conflict     &    0.018$^{***}$&  &  &  &   -0.061$^{*}$  \\
        &  (0.005) &  &  &  &  (0.033) \\
 - Political stability&    0.017$^{***}$&    0.029$^{***}$&    0.002$^{***}$&  &    0.097$^{*}$  \\
        &  (0.005) &  (0.001) &  (0.000) &  &  (0.048) \\
 - Population   &  &    0.394$^{***}$&    0.001$^{*}$  &    0.008$^{***}$&   -0.036$^{**}$ \\
        &  &  (0.076) &  (0.001) &  (0.000) &  (0.017) \\
 - Diaspora     &   -0.028$^{***}$&   -0.296$^{***}$&   -0.043$^{***}$&  &  \\
        &  (0.010) &  (0.024) &  (0.001) &  &  \\
 - Past migration&  &  &  &  &   -0.127$^{***}$\\
        &  &  &  &  &  (0.037) \\
 - Poverty      &   -0.015$^{**}$ &   -0.032$^{**}$ &   -0.001$^{***}$&  &  \\
        &  (0.006) &  (0.014) &  (0.000) &  &  \\
 - Geography    &  &   -0.095$^{***}$&   -0.006$^{***}$&  &  \\
        &  &  (0.021) &  (0.000) &  &  \\
 - Agriculture  &  &  &    0.002$^{*}$  &    0.008$^{***}$&  \\
        &  &  &  (0.001) &  (0.001) &  \\
 - Labor       &  &  &  &  &   -0.084$^{*}$  \\
        &  &  &  &  &  (0.047) \\
 - Urban        &  &  &  &   -0.016$^{***}$&  \\
        &  &  &  &  (0.000) &  \\
 - International aids&  &  &   -0.001$^{***}$&   -0.030$^{***}$&    0.107$^{**}$ \\
        &  &  &  (0.000) &  (0.004) &  (0.047) \\
        \midrule
\textit{Interacted terms (channels)}&  &  &  &  &  \\
 - Agriculture  &    0.005$^{**}$ &  &    0.007$^{***}$&   -0.005$^{***}$&   -0.027$^{***}$\\
        &  (0.002) &  &  (0.002) &  (0.001) &  (0.004) \\
 - International aid&   -0.031$^{***}$&  &  &  &   -0.039$^{***}$\\
        &  (0.005) &  &  &  &  (0.001) \\
 - Culture      &    0.019$^{**}$ &    0.015$^{***}$&  &  &    0.026$^{***}$\\
        &  (0.008) &  (0.001) &  &  &  (0.004) \\
 - Destination  &  &   -0.023$^{*}$  &  &  &  \\
        &  &  (0.011) &  &  &  \\
 - Diaspora     &  &  &    0.004$^{**}$ &  &  \\
        &  &  &  (0.001) &  &  \\
 - Poverty      &  &  &    0.004$^{***}$&    0.008$^{***}$&  \\
        &  &  &  (0.001) &  (0.000) &  \\
 - Education    &  &    0.034$^{***}$&  &  &  \\
        &  &  (0.001) &  &  &  \\
 - Environment  &  &  &  &    0.015$^{***}$&  \\
        &  &  &  &  (0.000) &  \\
 - Geography    &  &  &    0.025$^{***}$&  &  \\
        &  &  &  (0.001) &  &  \\
 - Income       &  &   -0.005$^{***}$&    0.010$^{***}$&  &   -0.014$^{***}$\\
        &  &  (0.000) &  (0.001) &  &  (0.001) \\
 - Past migration&    0.016$^{***}$&    0.014$^{***}$&    0.020$^{***}$&  &  \\
        &  (0.006) &  (0.001) &  (0.000) &  &  \\
 - Political stability&   -0.013$^{***}$&  &   -0.000$^{***}$&  &  \\
        &  (0.004) &  &  (0.000) &  &  \\
 - Urban        &  &  &  &    0.038$^{***}$&   -0.342$^{***}$\\
        &  &  &  &  (0.000) &  (0.026) \\
\bottomrule
\midrule
PEESE Correction: $\beta_0$&     0.047$^{**}$ &          -0.138$^{***}$&           2.938$^{***}$&          -0.464$^{***}$&           0.443$^{***}$\\
          &   [0.004,0.091]         & [-0.200,-0.077]         &   [2.637,3.238]         & [-0.468,-0.460]         &   [0.390,0.495]         \\
\midrule
N. of Obs     &     2,062 &      176 &      789 &      409 &      688 \\
N. of Studies     &     60 &      13 &      16 &      7 &      24 \\
\bottomrule
\multicolumn{6}{p{\dimexpr0.93\linewidth+3\tabcolsep}}{
{\bf Note}: {\linespread{0.5}\selectfont Step-wise regression of overall sample (1) and sub-samples defined by clusters (2-5) for fast-onset events. Estimates shown represent significant coefficients obtained through a step-wise procedure (not reported when not significant). Some coefficients and standard errors are zeros because of rounding. Controls are grouped by paper features, dependent variable, independent variable, sample and regression characteristics. PCC precision square weights ($1/se_i^2 $); robust standard errors clustered by study in parentheses; 95$\%$ confidence intervals in brackets; $^{*}$ \(p<0.10\), $^{**}$ \(p<0.05\), $^{***}$ \(p<0.01.\)  }}\\
\label{tab:MRA_fast_stepwise}
\end{longtable}
}}

The coefficient of $\hat{\beta}_0$, is positive and statistically significant in the overall sample and clusters 2 and 4, providing evidence of an increase in migration due to sudden natural hazards. It is worth noting that papers in Cluster 2 (column 3) mainly focus on fast-onset events and the summarized effect size is positive and very high. On the other side, the summarized effect of papers in clusters 1 and 3 is negative and statistically significant.

Results show evidence of publication bias for the overall sample and in Cluster 3, with $\hat{\beta}_1$ statistically significant signaling that the reported effect is not independent of its standard error. The significant and positive coefficient found for the published dummy confirms that there is a general \textit{Publication Impact}, so the peer-review process seems to affect the magnitude of the estimated effect, especially in clusters 1 and 2. Articles published in journals with higher \textit{Impact-factor} get higher estimates of the effects of natural disasters on migration, with exception of published articles in Cluster 2, suggesting that editors prefer to publish results that have a positive but more limited effect.
Natural disasters affect domestic and international migration flows. The positive coefficients of the group of controls related to the type of migration, in clusters 2 and 3 confirm that people respond to natural disasters with any kind of mobility. Specifically in Cluster 2 natural disasters increase both \textit{Internal} and \textit{Urbanization} migration, while studies in Cluster 3 find a greater effect on \textit{Internal} and \textit{International} movements of people. In Cluster 4, instead, estimates of the impact of natural disasters are lower in the case of \textit{Internal} migration. \textit{Hydrological} events have a greater impact on migration, the estimated coefficient is statistically significant in all clusters; if the fast-onset event refers to \textit{Geophysical}, \textit{Meteorological} and \textit{Climatological} disasters the effect on migration is lower.

The severity of natural disasters, such as hurricanes, landslides, or floods, affects regional agricultural production and it also has direct effects on employment and income in the agricultural sectors of the affected regions pushing people to migrate. However, on the one hand, natural disasters, such as droughts, floods, and storms, push individuals to move to find new sources of income or livelihood, on the other hand, natural disasters such as earthquakes, tsunamis, or hurricanes cause losses to populations that might lead people into a poverty trap, with potential migrants not having the resources to finance the trip. These effects, already highlighted by the literature, seem to be confirmed. Also in this literature, indeed, various controls and transmission channels analyzed in the original empirical models have a role in determining heterogeneity in results.

\section{Conclusions}
\label{sec: conclusions}
The present meta-analysis, aimed to systematically review and synthesize the empirical evidence on the relationship between environmental change and human migration, suggests that while there is a small, positive, and significant effect of slow- and rapid-onset environmental variables on migration, the heterogeneity of results in the existing economic literature highlights the need for a nuanced understanding of the causes and effects of environmental migration, as well as the specific characteristics of the places and populations involved.

If a key function of meta-analysis is to challenge and test the results of empirical studies, our study provides important insights that can inform both researchers and policymakers on the relationship between human migration and environmental changes or shocks.  Specifically, our findings suggest that a more nuanced and context-specific understanding of environmental migration is needed. Future research could profit from our work by exploring the average effect of specific environmental shocks, such as droughts or floods, and the important role of mediating factors that influence the decision to migrate, such as specific economic and social conditions.

The paper also offers an encompassing methodology for the empirical analysis of very heterogeneous outcomes of a research field. The sample collected through a systematic review of the literature, the bibliometric analysis, the construction of a co-citation network and the community detection on the structure of the network of essays, allow the inspection of a scientific area also in absence of a uniform and cohesive literature. In the case of environmental migration, the too many different characteristics in terms of object of analysis, empirical strategy, and mediating covariates render the meta-analytic average effect estimates just a first approximation of the quantitative evidence of the literature.

As shown in the present meta-analysis, when the level of heterogeneity in the outcome of a literature is relevant, as for the four clusters of papers that compose the economic literature on environmental migration, a group-by-group analysis has to be preferred and compared with the results of an overall meta-analysis.  

Moreover, our analysis highlights the need for greater collaboration and standardization of methods in the study of environmental migration. We report a lack of uniform and cohesive literature, with different studies using different methods, covariates, and definitions of key variables. This limits the external validity of existing results and calls for greater efforts by scholars and institutions to validate existing studies and improve the quality of data and methods used in future research.

Overall, our meta-analysis contributes to a better understanding of the complex relationship between slow or rapid environmental change and human migration. The implications of this work extend beyond the academic community to inform public policy and action. As environmen-tal change and human migration continue to characterize the global system, it is crucial for decision-makers to consider the insights provided by scientific research and for the scientific community to continue to produce results that improve the external validity of existing studies and help delineate evidence-based policies.

\section*{Competing Interests} 

The authors declare that there are no competing interests.

\bibliographystyle{apalike}
\bibliography{LCCNDM}

\end{document}